\documentstyle[onecolumn,epsf]{mn}

\newif\ifAMStwofonts

\newcommand{\sD}{{\cal D}}
\newcommand{\sF}{{\cal F}}
\newcommand{\sI}{{\cal I}}
\newcommand{\sV}{{\cal V}}

\ifoldfss
  \ifCUPmtlplainloaded \else
    \NewTextAlphabet{textbfit} {cmbxti10} {}
    \NewTextAlphabet{textbfss} {cmssbx10} {}
    \NewMathAlphabet{mathbfit} {cmbxti10} {} 
    \NewMathAlphabet{mathbfss} {cmssbx10} {} 
  \fi
  \ifAMStwofonts
    \ifCUPmtlplainloaded \else
      \NewSymbolFont{upmath} {eurm10}
      \NewSymbolFont{AMSa} {msam10}
      \NewMathSymbol{\upi}     {0}{upmath}{19}
      \NewMathSymbol{\umu}     {0}{upmath}{16}
      \NewMathSymbol{\upartial}{0}{upmath}{40}
      \NewMathSymbol{\leqslant}{3}{AMSa}{36}
      \NewMathSymbol{\geqslant}{3}{AMSa}{3E}

    \fi
  \fi
\fi 

\ifnfssone
  \newmathalphabet{\mathit}
  \addtoversion{normal}{\mathit}{cmr}{m}{it}
  \addtoversion{bold}{\mathit}{cmr}{bx}{it}
  \newmathalphabet{\mathbfit} 
  \addtoversion{normal}{\mathbfit}{cmr}{bx}{it}
  \addtoversion{bold}{\mathbfit}{cmr}{bx}{it}
  \newmathalphabet{\mathbfss} 
  \addtoversion{normal}{\mathbfss}{cmss}{bx}{n}
  \addtoversion{bold}{\mathbfss}{cmss}{bx}{n}
  \ifAMStwofonts
    \ifCUPmtlplainloaded \else
      \UseAMStwoboldmath
      \makeatletter
      \new@mathgroup\upmath@group
      \define@mathgroup\mv@normal\upmath@group{eur}{m}{n}
      \define@mathgroup\mv@bold\upmath@group{eur}{b}{n}
      \edef\UPM{\hexnumber\upmath@group}
      \new@mathgroup\amsa@group
      \define@mathgroup\mv@normal\amsa@group{msa}{m}{n}
      \define@mathgroup\mv@bold\amsa@group{msa}{m}{n}
      \edef\AMSa{\hexnumber\amsa@group}
      \makeatother
      \mathchardef\upi="0\UPM19
      \mathchardef\umu="0\UPM16
      \mathchardef\upartial="0\UPM40
      \mathchardef\leqslant="3\AMSa36
      \mathchardef\geqslant="3\AMSa3E
    \fi
  \fi
\fi 

\ifnfsstwo
  \DeclareMathAlphabet{\mathbfit}{OT1}{cmr}{bx}{it}
  \SetMathAlphabet\mathbfit{bold}{OT1}{cmr}{bx}{it}
  \DeclareMathAlphabet{\mathbfss}{OT1}{cmss}{bx}{n}
  \SetMathAlphabet\mathbfss{bold}{OT1}{cmss}{bx}{n}
  \ifAMStwofonts
    \ifCUPmtlplainloaded \else
      \DeclareSymbolFont{UPM}{U}{eur}{m}{n}
      \SetSymbolFont{UPM}{bold}{U}{eur}{b}{n}
      \DeclareSymbolFont{AMSa}{U}{msa}{m}{n}
      \DeclareMathSymbol{\upi}{0}{UPM}{"19}
      \DeclareMathSymbol{\umu}{0}{UPM}{"16}
      \DeclareMathSymbol{\upartial}{0}{UPM}{"40}
      \DeclareMathSymbol{\leqslant}{3}{AMSa}{"36}
      \DeclareMathSymbol{\geqslant}{3}{AMSa}{"3E}
    \fi
  \fi
\fi 

\ifCUPmtlplainloaded \else
  \ifAMStwofonts \else 
    \def\upi{\pi}
    \def\umu{\mu}
    \def\upartial{\partial}
  \fi
\fi

\title[The non-linear fluid dynamics of a warped accretion disc]{The
non-linear fluid dynamics of a warped accretion disc}
\author[G. I. Ogilvie]{G. I. Ogilvie$^{1,2,3}$\\
$^1$Institute of Astronomy, University of Cambridge, Madingley Road,
Cambridge CB3 0HA\\
$^2$Department of Applied Mathematics and Theoretical Physics,
University of Cambridge, Silver Street, Cambridge CB3 9EW\\
$^3$Isaac Newton Institute for Mathematical Sciences, University of
Cambridge, 20 Clarkson Road, Cambridge CB3 0EH}
\date{Accepted.  Received; in original form}

\pagerange{\pageref{firstpage}--\pageref{lastpage}}
\pubyear{1998}

\begin{document}

\maketitle

\label{firstpage}

\begin{abstract}
The dynamics of a viscous accretion disc subject to a slowly varying
warp of large amplitude is considered.  Attention is restricted to
discs in which self-gravitation is negligible, and to the generic case
in which the resonant wave propagation found in inviscid Keplerian
discs does not occur.  The equations of fluid dynamics are derived in
a coordinate system that follows the principal warping motion of the
disc.  They are reduced using asymptotic methods for thin discs, and
solved to extract the equation governing the warp.  In general, this
is a wave equation of parabolic type with non-linear dispersion and
diffusion, which describes fully non-linear bending waves.  This
method generalizes the linear theory of Papaloizou \& Pringle (1983)
to allow for an arbitrary rotation law, and extends it into the
non-linear domain, where it connects with a generalized version of the
theory of Pringle (1992).  The astrophysical implications of this
analysis are discussed briefly.
\end{abstract}


\section{Introduction}

There are many situations in astrophysics in which accretion discs are
believed to be non-planar, either because their profile is observed
directly (e.g. in the nucleus of the galaxy NGC 4258; Miyoshi
et~al. 1995) or in order to explain phenomena such as the periodic
modulation of light curves (e.g. in the X-ray binary Her X-1) or the
precession of jets (e.g. in the X-ray binary SS 433).  In many --
though not all -- applications the effects of self-gravitation can be
neglected and the evolution is dominated by viscous fluid dynamics.
The earliest attempts to derive equations governing the
time-dependence of a slowly varying warp of small amplitude (Petterson
1978; Hatchett, Begelman \& Sarazin 1981) suggested a simple diffusive
evolution but were shown by Papaloizou \& Pringle (1983) to be
incorrect at a fundamental level, being inconsistent with the
conservation of angular momentum.

Papaloizou \& Pringle (1983) demonstrated that the problem is
complicated because of the importance of a resonance in thin discs
that are both Keplerian (or very nearly so) and inviscid (or very
nearly so).  It is therefore necessary to distinguish between the {\it
generic (non-resonant) case\/} and the {\it resonant case\/} which
occurs when both
\begin{equation}
\left|{{\Omega^2-\kappa^2}\over{\Omega^2}}\right|\la
H/r\qquad\hbox{and}\qquad\alpha\la H/r.
\end{equation}
Here $\Omega$ is the angular velocity, $\kappa$ is the epicyclic
frequency, $H/r$ is the ratio of the semi-thickness of the disc to the
radius, and $\alpha$ is the dimensionless viscosity parameter (Shakura
\& Sunyaev 1973).  Unlike most resonances in discs, which occur at
specific radii, this one, which is effectively the $m=1$ inner
Lindblad resonance, is likely to occur either everywhere in the disc
or nowhere.  Current estimates suggest that the resonant case is not
relevant to discs in X-ray binaries or active galactic nuclei, but may
apply in protostellar discs.  However, the resonance is delicate and
could be destroyed by the effects of self-gravitation, magnetic
fields, or turbulence.

Papaloizou \& Pringle (1983) derived the first consistent equation
governing a warp of small amplitude.  They considered the
(non-resonant) case of a Keplerian disc with a significant viscosity,
treating the warp as a slowly varying disturbance with azimuthal wave
number $m=1$, using linear Eulerian perturbation theory.  The
resulting equation for the warp is a complex linear diffusion
equation, which has been used in applications (e.g. Kumar \& Pringle
1985).

Papaloizou \& Lin (1994)\footnote{See also Papaloizou \& Lin (1995),
in which self-gravitation is included.} considered the case of an
inviscid disc.  Again, linear Eulerian perturbation theory was used,
and the authors assumed the warp to be a slowly varying normal mode of
the disc.  In the non-Keplerian (non-resonant) case the warp obeys a
dispersive linear wave equation, while in the Keplerian (resonant)
case the warp obeys a non-dispersive linear wave equation.  By
considering the effect of a small viscosity on the inviscid modes,
Papaloizou \& Lin (1994) connected this theory with that of Papaloizou
\& Pringle (1983), showing how the transition occurs between wave-like
and diffusive behaviour in Keplerian discs when $\alpha\approx H/r$.
This theory has also been used in applications (e.g. Papaloizou \&
Terquem 1995).

The major uncertainty in these theories is whether the linear analysis
is valid for warps of a sufficient amplitude to be observable.  The
linear Eulerian perturbation theory is formally valid only when the
tilt angle $\beta(r,t)$ of the warp satisfies $|\beta|\ll H/r$.
However, because of the special nature of the mode involved, which
differs little locally from a rigid tilt $\beta={\rm constant}$, one
suspects that the linear theory might be valid for larger warps.  In
fact, a more appropriate measure of the amplitude of the warp is
$|\partial\beta/\partial\ln r|$.  The Eulerian perturbation theory
offers little information about possible non-linear effects, although
it does predict that, in the resonant case, an amplitude
$|\partial\beta/\partial\ln r|\approx H/r$ would result in horizontal
shearing motions in the disc comparable to the sound speed, which are
expected to be unstable (Kumar \& Coleman 1993; Gammie, Goodman \&
Ogilvie, in preparation).

Meanwhile, Pringle (1992) developed a different approach in which the
forms of the equations governing a warped viscous disc are derived
simply by requiring mass and angular momentum to be conserved, but
without reference to the detailed internal fluid dynamics of the disc.
In this scheme, neighbouring rings in the disc exchange angular
momentum by means of viscous torques which are of two kinds.  One kind
of torque (associated with a kinematic viscosity coefficient $\nu_1$)
acts on the differential rotation in the plane of the disc, and leads
to accretion; the other kind (associated with $\nu_2$) acts to flatten
the disc.  The equations derived are
\begin{equation}
{{\partial\Sigma}\over{\partial t}}+{{1}\over{r}}{{\partial}\over
{\partial r}}(r\Sigma\bar v_r)=0\label{jep1}
\end{equation}
for the surface density $\Sigma(r,t)$, and
\begin{equation}
{{\partial}\over{\partial t}}(\Sigma
r^2\Omega{\bmath\ell})+{{1}\over{r}}{{\partial}\over{\partial
r}}(\Sigma\bar
v_rr^3\Omega{\bmath\ell})={{1}\over{r}}{{\partial}\over{\partial
r}}\left(\nu_1\Sigma r^3{{{\rm d}\Omega}\over{{\rm
d}r}}{\bmath\ell}\right)+{{1}\over{r}}{{\partial}\over{\partial
r}}\left({\textstyle{{1}\over{2}}}\nu_2\Sigma
r^3\Omega{{\partial{\bmath\ell}}\over{\partial r}}\right)\label{jep2}
\end{equation}
for the angular momentum, in the absence of external torques.  Here
$\bar v_r(r,t)$ is the mean radial velocity and ${\bmath\ell}(r,t)$ is
the tilt vector, which is a unit vector parallel to the orbital
angular momentum of the ring.  From this follow the equations
\begin{equation}
\Sigma\bar v_r{{{\rm d}(r^2\Omega)}\over{{\rm
d}r}}={{1}\over{r}}{{\partial}\over{\partial r}}\left(\nu_1\Sigma
r^3{{{\rm d}\Omega}\over{{\rm
d}r}}\right)-{\textstyle{{1}\over{2}}}\nu_2\Sigma
r^2\Omega\left|{{\partial{\bmath\ell}}\over{\partial
r}}\right|^2\label{jep3}
\end{equation}
for the component of angular momentum parallel to ${\bmath\ell}$, and
\begin{equation}
\Sigma r^2\Omega\left[{{\partial{\bmath\ell}}\over{\partial
t}}+\left(\bar v_r-\nu_1{{{\rm d}\ln\Omega}\over{{\rm
d}r}}\right){{\partial{\bmath\ell}}\over{\partial
r}}\right]={{1}\over{r}}{{\partial}\over{\partial
r}}\left({\textstyle{{1}\over{2}}}\nu_2\Sigma
r^3\Omega{{\partial{\bmath\ell}}\over{\partial
r}}\right)+{\textstyle{{1}\over{2}}}\nu_2\Sigma
r^2\Omega\left|{{\partial{\bmath\ell}}\over{\partial
r}}\right|^2{\bmath\ell}\label{jep4}
\end{equation}
for the tilt vector.

This approach appears to be valid for warps of large amplitude and
would therefore represent an advance on the linear theory.  It has
been used in several applications, in particular to identify the
radiation-driven instability (Pringle 1996) and to explore its linear
theory (e.g. Maloney, Begelman \& Pringle 1996) and non-linear
evolution (e.g. Pringle 1997).  However, because the equations of
Pringle (1992) are derived somewhat heuristically, without reference
to the detailed internal fluid dynamics of the disc, some doubts
remain over the validity of this method.  The obvious questions to be
addressed are whether any internal degree of freedom of the rings has
been neglected, whether the interaction between neighbouring rings is
purely of the assumed form of viscous torques, and whether there are
any non-linear fluid-dynamical effects that might limit the amplitude
of the warp.  It is therefore important to investigate whether the
equations of Pringle (1992) can be derived {\it ab initio\/} from the
three-dimensional fluid-dynamical equations, and to understand how
they connect to the previously established linear theory.  These are
the aims of this paper.

The analysis is organized as follows.  I first define a coordinate
system that follows the principal warping motion of the disc, and
derive the basic results necessary for vector calculus (Section~2).  I
then present the exact forms of the equations of fluid dynamics in
this coordinate system (Section~3).  The equations are reduced using
an asymptotic analysis for thin discs (Section~4).  I then apply
separation of variables to solve the fully non-linear problem except
for the determination of three dimensionless coefficients from the
solution of a set of ordinary differential equations (ODEs)
(Section~5).  These are evaluated in Section~6.  Readers interested
only in the interpretation and application of this work may proceed to
Section~7, where the principal results are summarized and discussed
further.

\section{Definition of warped coordinates}

The non-linear fluid dynamics of a warped disc is most naturally
described in the case of a spherically symmetric external potential.
Then a flat disc has no preferred plane of orientation, and must
possess a zero-frequency mode consisting of a rigid tilt of any
amplitude.  Continuous with this mode, there exist bending waves, with
azimuthal wave number $m=1$ in linear theory, which vary on a
time-scale long compared to the orbital time-scale, and on a
length-scale long compared to the thickness of the disc.  This is in
contrast to the other modes of the disc (Lubow \& Pringle 1993) and
motivates the following analysis.

\subsection{Coordinates and basis vectors}

Let $(x,y,z)$ be Cartesian coordinates in an inertial frame of
reference, and define the spherical radial coordinate
$r=(x^2+y^2+z^2)^{1/2}$.  On each sphere $r={\rm constant}$, define
the usual angular coordinates $(\theta,\phi)$, but with respect to an
axis that is tilted to point in the direction of the unit vector
${\bmath\ell}(r,t)$ (Fig.~1).  It is intended that the disc matter on
each sphere will lie close to $\theta=\pi/2$.  The tilt vector can be
described by Euler angles $\beta(r,t)$ and $\gamma(r,t)$:
\begin{equation}
{\bmath\ell}=\sin\beta\cos\gamma\,{\bmath
e}_x+\sin\beta\sin\gamma\,{\bmath e}_y+\cos\beta\,{\bmath e}_z.
\end{equation}
In detail, warped spherical polar coordinates $(r,\theta,\phi)$ are
defined by
\begin{equation}
\left[\matrix{x\cr y\cr z\cr}\right]={\bf
M}(r,\theta,\phi,t)\left[\matrix{r\cr 0\cr 0\cr}\right],
\end{equation}
where
\begin{equation}
{\bf M}(r,\theta,\phi,t)=\left[\matrix{\cos\gamma&-\sin\gamma&0\cr
\sin\gamma&\cos\gamma&0\cr
0&0&1\cr}\right]\left[\matrix{\cos\beta&0&\sin\beta\cr 0&1&0\cr
-\sin\beta&0&\cos\beta\cr}\right]\left[\matrix{\cos\phi&-\sin\phi&0\cr
\sin\phi&\cos\phi&0\cr
0&0&1\cr}\right]\left[\matrix{\sin\theta&\cos\theta&0\cr 0&0&1\cr
\cos\theta&-\sin\theta&0\cr}\right]\label{m1m2m3m4}
\end{equation}
is a composition of four orthogonal linear transformations.  For the
functions $\beta(r,t)$ and $\gamma(r,t)$, a prime and a dot will
denote differentiation with respect to $r$ and $t$, respectively.

\begin{figure}
\centerline{\epsfbox{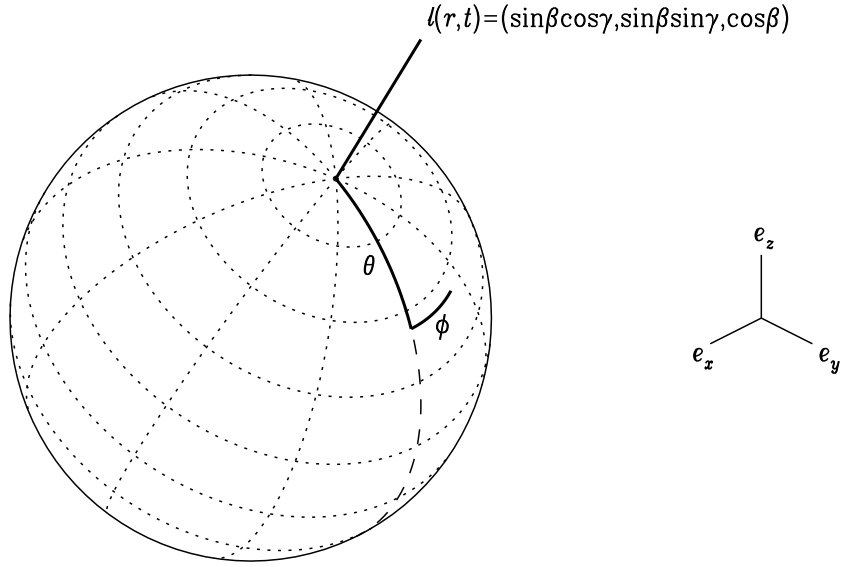}}
\caption{Warped spherical polar coordinates $(r,\theta,\phi)$.  The
axis of each sphere $r={\rm constant}$ is tilted to point in the
direction of the unit vector ${\bmath\ell}(r,t)$.}
\end{figure}

Although $(r,\theta,\phi)$ are not orthogonal coordinates, it is
appropriate to refer vectors to a natural orthonormal basis $\{{\bmath
e}_r,{\bmath e}_\theta,{\bmath e}_\phi\}$ defined by
\begin{equation}
\left[\matrix{{\bmath e}_x\cr {\bmath e}_y\cr {\bmath
e}_z\cr}\right]={\bf M}\left[\matrix{{\bmath e}_r\cr {\bmath
e}_\theta\cr {\bmath e}_\phi\cr}\right].
\end{equation}
The vector ${\bmath e}_r$ is normal to the sphere $r={\rm constant}$,
while the vectors ${\bmath e}_\theta$ and ${\bmath e}_\phi$ are
tangent to it.  The components of an arbitrary vector ${\bmath F}$
then transform according to
\begin{equation}
\left[\matrix{F_x\cr F_y\cr F_z\cr}\right]={\bf M}\left[\matrix{F_r\cr
F_\theta\cr F_\phi\cr}\right].
\end{equation}

\subsection{Calculus in warped coordinates}

The matrix ${\bf M}$ is orthogonal.  Its decomposition in equation
(\ref{m1m2m3m4}) makes it easy to compute the differential
\begin{equation}
{\rm d}{\bf M}={\bf M}{\bf A},
\end{equation}
where ${\bf A}$ is an antisymmetric matrix, with components given by
\begin{eqnarray}
A_{12}&=&-{\rm d}\theta-\cos\phi\,{\rm d}\beta-\sin\beta\sin\phi\,{\rm
d}\gamma,\\ A_{13}&=&-\sin\theta\,{\rm d}\phi+\cos\theta\sin\phi\,{\rm
d}\beta-(\cos\beta\sin\theta+\sin\beta\cos\theta\cos\phi)\,{\rm
d}\gamma,\\ A_{23}&=&-\cos\theta\,{\rm d}\phi-\sin\theta\sin\phi\,{\rm
d}\beta-(\cos\beta\cos\theta-\sin\beta\sin\theta\cos\phi)\,{\rm
d}\gamma.
\end{eqnarray}
The coordinate differentials are therefore related by
\begin{equation}
\left[\matrix{{\rm d}x\cr {\rm d}y\cr {\rm d}z\cr}\right]={\bf
M}\left[\matrix{{\rm d}r\cr r\,{\rm d}\theta+r\cos\phi\,{\rm
d}\beta+r\sin\beta\sin\phi\,{\rm d}\gamma\cr r\sin\theta\,{\rm
d}\phi-r\cos\theta\sin\phi\,{\rm
d}\beta+r(\cos\beta\sin\theta+\sin\beta\cos\theta\cos\phi)\,{\rm
d}\gamma\cr}\right],
\end{equation}
so the Jacobian matrix of the transformation is
\begin{equation}
{\bf J}=\left[\matrix{\partial x/\partial r&\partial
x/\partial\theta&\partial x/\partial\phi\cr \partial y/\partial
r&\partial y/\partial\theta&\partial y/\partial\phi\cr \partial
z/\partial r&\partial z/\partial\theta&\partial
z/\partial\phi\cr}\right]={\bf M}\left[\matrix{1&0&0\cr
r\beta^\prime\cos\phi+r\gamma^\prime\sin\beta\sin\phi&r&0\cr
-r\beta^\prime\cos\theta\sin\phi+r\gamma^\prime(\cos\beta\sin\theta+
\sin\beta\cos\theta\cos\phi)&0&r\sin\theta\cr}\right],
\end{equation}
with determinant
\begin{equation}
J=r^2\sin\theta,
\end{equation}
as for ordinary spherical polar coordinates.  

The gradient operator takes the form
\begin{equation}
\nabla={\bmath e}_r\sD+{\bmath
e}_\theta\,{{1}\over{r}}\partial_\theta+{\bmath
e}_\phi\,{{1}\over{r\sin\theta}}\partial_\phi,
\end{equation}
where
\begin{equation}
\sD=\partial_r-(\beta^\prime\cos\phi+\gamma^\prime\sin\beta\sin\phi)
\partial_\theta-\left[-\beta^\prime\cos\theta\sin\phi+\gamma^\prime
(\cos\beta\sin\theta+\sin\beta\cos\theta\cos\phi)\right]
{{1}\over{\sin\theta}}\partial_\phi
\end{equation}
is a modified radial derivative.

Finally, the condition
\begin{equation}
\left[\matrix{{\rm d}{\bmath e}_x\cr {\rm d}{\bmath e}_y\cr {\rm
d}{\bmath e}_z\cr}\right]={\bf0}
\end{equation}
yields the differentials of the unit vectors in the form
\begin{eqnarray}
{\rm d}{\bmath e}_r&=&{\bmath e}_\theta({\rm d}\theta+\cos\phi\,{\rm
d}\beta+\sin\beta\sin\phi\,{\rm d}\gamma)+{\bmath
e}_\phi\left[\sin\theta\,{\rm d}\phi-\cos\theta\sin\phi\,{\rm
d}\beta+(\cos\beta\sin\theta+\sin\beta\cos\theta\cos\phi)\,{\rm
d}\gamma\right],\\ {\rm d}{\bmath e}_\theta&=&-{\bmath e}_r({\rm
d}\theta+\cos\phi\,{\rm d}\beta+\sin\beta\sin\phi\,{\rm
d}\gamma)+{\bmath e}_\phi\left[\cos\theta\,{\rm
d}\phi+\sin\theta\sin\phi\,{\rm
d}\beta+(\cos\beta\cos\theta-\sin\beta\sin\theta\cos\phi)\,{\rm
d}\gamma\right],\label{detheta}\\ {\rm d}{\bmath e}_\phi&=&-{\bmath
e}_r\left[\sin\theta\,{\rm d}\phi-\cos\theta\sin\phi\,{\rm
d}\beta+(\cos\beta\sin\theta+\sin\beta\cos\theta\cos\phi)\,{\rm
d}\gamma\right]\nonumber\\ &&\qquad-{\bmath
e}_\theta\left[\cos\theta\,{\rm d}\phi+\sin\theta\sin\phi\,{\rm
d}\beta+(\cos\beta\cos\theta-\sin\beta\sin\theta\cos\phi)\,{\rm
d}\gamma\right].
\end{eqnarray}
In particular,
\begin{eqnarray}
\nabla{\bmath e}_r&=&{{1}\over{r}}\,{\bmath e}_\theta{\bmath
e}_\theta+{{1}\over{r}}\,{\bmath e}_\phi{\bmath e}_\phi,\\
\nabla{\bmath e}_\theta&=&-{{1}\over{r}}\,{\bmath e}_\theta{\bmath
e}_r+{{\cot\theta}\over{r}}\,{\bmath e}_\phi{\bmath
e}_\phi+{{f}\over{r\sin\theta}}\,{\bmath e}_r{\bmath e}_\phi,\\
\nabla{\bmath e}_\phi&=&-{{1}\over{r}}\,{\bmath e}_\phi{\bmath
e}_r-{{\cot\theta}\over{r}}\,{\bmath e}_\phi{\bmath
e}_\theta-{{f}\over{r\sin\theta}}\,{\bmath e}_r{\bmath e}_\theta,
\end{eqnarray}
where
\begin{equation}
f=r\beta^\prime\sin\phi-r\gamma^\prime\sin\beta\cos\phi.\label{f}
\end{equation}

The advantages of this method over that of Petterson (1978) are to be
emphasized.  Petterson (1978) introduced a warped coordinate system
based on tilted cylinders, rather than spheres.  This is less
satisfactory because the resulting coordinate system has a non-trivial
Jacobian determinant and, in fact, breaks down for large-amplitude
warps.  The present method does not suffer from these defects.
Moreover, it can be understood using elementary vector calculus rather
than requiring advanced differential geometry.

\subsection{Absolute and relative velocities}

Consider a fluid element with coordinates $x(t)$, $y(t)$ and $z(t)$.
The velocity components are
\begin{equation}
\left[\matrix{u_x\cr u_y\cr u_z\cr}\right]=\left[\matrix{{\rm d}x/{\rm
d}t\cr {\rm d}y/{\rm d}t\cr {\rm d}z/{\rm d}t\cr}\right]={\bf
M}\left[\matrix{u_r\cr u_\theta\cr u_\phi\cr}\right],
\end{equation}
where
\begin{equation}
\left[\matrix{u_r\cr u_\theta\cr u_\phi\cr}\right]=\left[\matrix{{\rm
d}r/{\rm d}t\cr r({\rm d}\theta/{\rm d}t)+r({\rm d}\beta/{\rm
d}t)\cos\phi+r({\rm d}\gamma/{\rm d}t)\sin\beta\sin\phi\cr
r\sin\theta({\rm d}\phi/{\rm d}t)-r({\rm d}\beta/{\rm
d}t)\cos\theta\sin\phi+r({\rm d}\gamma/{\rm
d}t)(\cos\beta\sin\theta+\sin\beta\cos\theta\cos\phi)\cr}\right].
\end{equation}
Here, ${\rm d}/{\rm d}t$ stands for the Lagrangian time derivative
\begin{equation}
{\rm D}=(\partial_t)_{x,y,z}+u_x\partial_x+u_y\partial_y+u_z\partial_z.
\end{equation}
The components $(u_r,u_\theta,u_\phi)$ are the {\it absolute} velocity
components, in the sense that ${\bmath u}=u_r\,{\bmath
e}_r+u_\theta\,{\bmath e}_\theta+u_\phi\,{\bmath e}_\phi$ is the
velocity vector as measured in the inertial frame.  It is intended,
however, that the warped coordinate system itself will follow the
principal motion of the fluid, other than its orbital angular
velocity.  The additional motion with respect to the warped coordinate
system is described by the {\it relative} velocity components
\begin{equation}
\left[\matrix{v_r\cr v_\theta\cr v_\phi\cr}\right]=\left[\matrix{{\rm
d}r/{\rm d}t\cr r({\rm d}\theta/{\rm d}t)\cr r\sin\theta({\rm
d}\phi/{\rm d}t)\cr}\right].
\end{equation}
The Lagrangian time derivative takes the form
\begin{equation}
{\rm
D}=(\partial_t)_{r,\theta,\phi}+v_r\partial_r+{{v_\theta}\over{r}}
\partial_\theta+{{v_\phi}\over{r\sin\theta}}\partial_\phi.\label{d}
\end{equation}

The absolute velocity may then be written
\begin{equation}
{\bmath u}=u_r\,{\bmath e}_r+u_\theta\,{\bmath
e}_\theta+u_\phi\,{\bmath e}_\phi,
\end{equation}
with
\begin{eqnarray}
u_r&=&v_r,\label{ur}\\ u_\theta&=&v_\theta+r({\rm
D}\beta)\cos\phi+r({\rm D}\gamma)\sin\beta\sin\phi,\label{utheta}\\
u_\phi&=&v_\phi-r({\rm D}\beta)\cos\theta\sin\phi+r({\rm
D}\gamma)(\cos\beta\sin\theta+\sin\beta\cos\theta\cos\phi).\label{uphi}
\end{eqnarray}

\section{Fluid dynamics in warped coordinates}

\subsection{Basic equations}

The equations governing the dynamics of a compressible,
non-self-gravitating fluid are the equation of mass conservation,
\begin{equation}
{\rm D}\rho=-\rho\nabla\!\cdot\!{\bmath u},\label{drho}
\end{equation}
the adiabatic condition,
\begin{equation}
{\rm D}p=-\Gamma p\nabla\!\cdot\!{\bmath u},\label{dp}
\end{equation}
and the equation of motion,
\begin{equation}
\rho{\rm D}{\bmath u}=-\nabla
p-\rho\nabla\Phi+\nabla\!\cdot\!\left[\mu\nabla{\bmath
u}+\mu(\nabla{\bmath u})^{\rm T}\right]+\nabla\left[(\mu_{\rm
b}-{\textstyle{{2}\over{3}}}\mu)\nabla\!\cdot\!{\bmath
u}\right]+{\bmath F}.\label{du}
\end{equation}
Here $\rho$ is the density, $p$ is the pressure, $\Gamma$ is the
adiabatic exponent, $\Phi$ is the external gravitational potential,
$\mu$ and $\mu_{\rm b}$ are the shear and bulk viscosities, and
${\bmath F}$ is an arbitrary external force.

In using these equations, it is assumed that the agent responsible for
angular momentum transport in accretion discs can be treated as an
isotropic viscosity in the sense of the Navier-Stokes equation.  In
the case of magnetohydrodynamic turbulence, it is not known how
accurate this assumption is, nor whether any alternative constitutive
relation for the turbulent stress would be better.  Although bulk
viscosity is not usually discussed in the context of accretion discs,
it is a standard component in the compressible Navier-Stokes equation
and is included here for completeness.  It is, however, assumed that
the viscosities are only dynamically, rather than thermodynamically,
important.  A more sophisticated treatment would include the thermal
effects of viscous dissipation and radiative transport.

Using the information given in Section~2, it is possible to derive the
form of these equations in warped coordinates.  Either absolute or
relative velocity components are used below, whichever gives the most
compact form of the equations.  First, note that
$\nabla\!\cdot\!{\bmath u}$ may be written in two equivalent ways:
\begin{eqnarray}
\nabla\!\cdot\!{\bmath u}&=&{{1}\over{r^2}}\sD(r^2
u_r)+{{1}\over{r\sin\theta}}\partial_\theta(u_\theta\sin\theta)+
{{1}\over{r\sin\theta}}\partial_\phi u_\phi\nonumber\\
&=&{{1}\over{r^2}}\partial_r(r^2v_r)+{{1}\over{r\sin\theta}}
\partial_\theta(v_\theta\sin\theta)+{{1}\over{r\sin\theta}}\partial_\phi
v_\phi.
\end{eqnarray}
The familiar form of the latter expression is a consequence of the
fact that the Jacobian determinant of the warped coordinate system is
equal to that of ordinary spherical polar coordinates.  This deals
with equations (\ref{drho}) and (\ref{dp}).  Equation (\ref{du})
requires a lengthy calculation, especially to evaluate the viscous
terms, although much of their complexity is already present in
ordinary spherical polar coordinates.  The result is
\begin{eqnarray}
\lefteqn{\rho\left({\rm
D}u_r-{{u_\theta^2}\over{r}}-{{u_\phi^2}\over{r}}\right)=-\sD
p-\rho\sD\Phi+F_r}&\nonumber\\ &&+\sD\left[(\mu_{\rm
b}+{\textstyle{{1}\over{3}}}\mu)\nabla\!\cdot\!{\bmath
u}\right]+{{1}\over{r^2}}\sD(\mu r^2\sD
u_r)+{{1}\over{r^2\sin\theta}}\partial_\theta(\mu\sin\theta\,\partial_\theta
u_r)+{{1}\over{r^2\sin^2\theta}}\partial_\phi(\mu\partial_\phi
u_r)-{{2u_r}\over{r^2}}\sD(\mu r)\nonumber\\ &&-{{\sD(\mu
r^2)}\over{r^3\sin\theta}}\partial_\theta(u_\theta\sin\theta)+
(\partial_\theta\mu)\sD\left({{u_\theta}\over{r}}\right)-{{\sD(\mu
r^2)}\over{r^3\sin\theta}}\partial_\phi
u_\phi+{{\partial_\phi\mu}\over{\sin\theta}}\sD\left({{u_\phi}\over{r}}
\right)+{{f}\over{r^2\sin\theta}}\left({{u_\theta}\over{\sin\theta}}
\partial_\phi\mu-u_\phi\partial_\theta\mu\right),\label{dur}
\end{eqnarray}
\begin{eqnarray}
\lefteqn{\rho\left\{{\rm D}u_\theta+{{u_ru_\theta}\over{r}}-
{{u_\phi}\over{r\sin\theta}}\left[u_\phi\cos\theta+r({\rm
D}\beta)\sin\phi-r({\rm
D}\gamma)\sin\beta\cos\phi\right]\right\}=-{{1}\over{r}}\partial_\theta
p-{{\rho}\over{r}}\partial_\theta\Phi+F_\theta}&\nonumber\\
&&+{{1}\over{r}}\partial_\theta\left[(\mu_{\rm
b}+{\textstyle{{1}\over{3}}}\mu)\nabla\!\cdot\!{\bmath
u}\right]+{{1}\over{r^2}}\sD(\mu r^2\sD
u_\theta)+{{1}\over{r^2\sin\theta}}\partial_\theta(\mu\sin\theta\,
\partial_\theta
u_\theta)+{{1}\over{r^2\sin^2\theta}}\partial_\phi(\mu\partial_\phi
u_\theta)\nonumber\\ &&-{{u_\theta}\over{r^2}}\sD(\mu
r)-{{u_\theta\cos\theta}\over{r^2\sin^2\theta}}\partial_\theta
(\mu\sin\theta)+{{\sD(\mu r^2)}\over{r^3}}\partial_\theta
u_r-{{\partial_\theta\mu}\over{r}}\sD
u_r-{{\partial_\theta(\mu\sin^2\theta)}\over{r^2\sin^3\theta}}\partial_\phi
u_\phi+{{\partial_\phi\mu}\over{r^2}}\partial_\theta\left(
{{u_\phi}\over{\sin\theta}}\right)\nonumber\\
&&-{{1}\over{r^2}}\sD\left({{\mu
rfu_\phi}\over{\sin\theta}}\right)-{{\mu f}\over{r\sin\theta}}\sD
u_\phi-{{\mu f^2u_\theta}\over{r^2\sin^2\theta}},\label{dutheta}
\end{eqnarray}
\begin{eqnarray}
\lefteqn{\rho\left\{{\rm
D}u_\phi+{{u_ru_\phi}\over{r}}+{{u_\theta}\over{r\sin\theta}}
\left[u_\phi\cos\theta+r({\rm D}\beta)\sin\phi-r({\rm
D}\gamma)\sin\beta\cos\phi\right]\right\}=-{{1}\over{r\sin\theta}}
\partial_\phi
p-{{\rho}\over{r\sin\theta}}\partial_\phi\Phi+F_\phi}&\nonumber\\
&&+{{1}\over{r\sin\theta}}\partial_\phi\left[(\mu_{\rm
b}+{\textstyle{{1}\over{3}}}\mu)\nabla\!\cdot\!{\bmath
u}\right]+{{1}\over{r^2}}\sD(\mu r^2\sD
u_\phi)+{{1}\over{r^2\sin\theta}}\partial_\theta(\mu\sin\theta\,
\partial_\theta
u_\phi)+{{1}\over{r^2\sin^2\theta}}\partial_\phi(\mu\partial_\phi
u_\phi)\nonumber\\ &&-{{u_\phi}\over{r^2}}\sD(\mu
r)-{{u_\phi\cos\theta}\over{r^2\sin^2\theta}}\partial_\theta
(\mu\sin\theta)+{{\sD(\mu r^2)}\over{r^3\sin\theta}}\partial_\phi
u_r-{{\partial_\phi\mu}\over{r\sin\theta}}\sD
u_r+{{\partial_\theta(\mu\sin^2\theta)}\over{r^2\sin^3\theta}}
\partial_\phi u_\theta-{{\partial_\phi\mu}\over{r^2}}\partial_\theta
\left({{u_\theta}\over{\sin\theta}}\right)\nonumber\\
&&+{{1}\over{r^2}}\sD\left({{\mu
rfu_\theta}\over{\sin\theta}}\right)+{{\mu f}\over{r\sin\theta}}\sD
u_\theta-{{\mu f^2u_\phi}\over{r^2\sin^2\theta}}.\label{duphi}
\end{eqnarray}
These equations must be solved in conjunction with equations
(\ref{ur})--(\ref{uphi}) defining the absolute velocity components and
equation (\ref{d}) defining the Lagrangian time derivative.

\subsection{Generalized form of the angular momentum equation}

The equations above are capable of describing any flow, with
$\beta(r,t)$ and $\gamma(r,t)$ being arbitrarily prescribed
(differentiable) functions.  In order to derive meaningful dynamical
equations for $\beta$ and $\gamma$, one must impose suitable
constraints on the flow.  For a thin disc, the fluid must remain close
to the surface $\theta=\pi/2$, and this condition is enforced by the
asymptotic analysis presented in Section~4.  It will be convenient to
work with the dimensionless complex variable
\begin{equation}
\psi=\psi_{\rm r}+{\rm i}\psi_{\rm i}=r(\beta^\prime+{\rm
i}\gamma^\prime\sin\beta).
\end{equation}
Then $|\psi|$ is a measure of the amplitude of the warp, with
\begin{equation}
|\psi|=r\left|{{\partial{\bmath\ell}}\over{\partial r}}\right|.
\end{equation}

It will be found that, although equation (\ref{jep1}) for the surface
density can easily be derived from the equations of this section in
the limit of a thin disc, it is impossible to obtain equations
(\ref{jep3}) and (\ref{jep4}) in a similar manner.  One must allow for
a more general form of the torque between neighbouring rings, and
attempt to derive an angular momentum equation of the form
\begin{equation}
{{\partial}\over{\partial t}}(\Sigma
r^2\Omega{\bmath\ell})+{{1}\over{r}}{{\partial}\over{\partial
r}}(\Sigma\bar
v_rr^3\Omega{\bmath\ell})={{1}\over{r}}{{\partial}\over{\partial
r}}\left(Q_1\sI
r^2\Omega^2{\bmath\ell}\right)+{{1}\over{r}}{{\partial}\over{\partial
r}}\left(Q_2\sI r^3\Omega^2{{\partial{\bmath\ell}}\over{\partial
r}}\right)+{{1}\over{r}}{{\partial}\over{\partial r}}\left(Q_3\sI
r^3\Omega^2{\bmath\ell}\times{{\partial{\bmath\ell}}\over{\partial
r}}\right).
\end{equation}
Here $\sI(r,t)$ is the azimuthally averaged second vertical moment of
the density, which in cylindrical polar coordinates would be written
\begin{equation}
\sI={{1}\over{2\pi}}\int_0^{2\pi}\tilde{\sI}\,{\rm
d}\phi,\qquad\hbox{where}\quad\tilde{\sI}=\int_{-\infty}^\infty\rho
z^2\,{\rm d}z,\label{i}
\end{equation}
and is an important dynamical quantity in the theory of bending waves.
The coefficients $Q_1$, $Q_2$ and $Q_3$ are dimensionless quantities
to be determined, which will be found to depend on the rotation law
and the shear viscosity, and also, in the non-linear theory, on the
adiabatic exponent, the bulk viscosity and the amplitude of the warp.
From this follow the equations
\begin{equation}
\Sigma\bar
v_r(r^2\Omega)^\prime={{1}\over{r}}{{\partial}\over{\partial
r}}\left(Q_1\sI r^2\Omega^2\right)-Q_2\sI
r^2\Omega^2\left|{{\partial{\bmath\ell}}\over{\partial r}}\right|^2
\end{equation}
for the component of angular momentum parallel to ${\bmath\ell}$, and
\begin{equation}
\Sigma r^2\Omega\left({{\partial{\bmath\ell}}\over{\partial t}}+\bar
v_r{{\partial{\bmath\ell}}\over{\partial r}}\right)=Q_1\sI
r\Omega^2{{\partial{\bmath\ell}}\over{\partial
r}}+{{1}\over{r}}{{\partial}\over{\partial r}}\left(Q_2\sI
r^3\Omega^2{{\partial{\bmath\ell}}\over{\partial r}}\right)+Q_2\sI
r^2\Omega^2\left|{{\partial{\bmath\ell}}\over{\partial
r}}\right|^2{\bmath\ell}+{{1}\over{r}}{{\partial}\over{\partial
r}}\left(Q_3\sI
r^3\Omega^2{\bmath\ell}\times{{\partial{\bmath\ell}}\over{\partial
r}}\right)
\end{equation}
for the tilt vector.  The correspondence with $\beta$ and $\gamma$ is
made through the relation
\begin{equation}
{\bmath\ell}=-{\bmath e}_\theta\big|_{\theta=\pi/2},
\end{equation}
which can be differentiated using equation (\ref{detheta}).  Thus
\begin{equation}
\Sigma\bar v_r(r^2\Omega)^\prime={{1}\over{r}}\partial_r(Q_1\sI
r^2\Omega^2)-Q_2\sI\Omega^2|\psi|^2,\label{new1}
\end{equation}
\begin{eqnarray}
\Sigma r^2\Omega(\dot\beta+\bar v_r\beta^\prime)&=&X\nonumber\\
&=&Q_1\sI r\Omega^2\beta^\prime+{{1}\over{r}}\partial_r\left[\sI
r^3\Omega^2(Q_2\beta^\prime-Q_3\gamma^\prime\sin\beta)\right]-\sI
r^2\Omega^2\gamma^\prime\cos\beta(Q_2\gamma^\prime\sin\beta+
Q_3\beta^\prime),\label{new2}
\end{eqnarray}
and
\begin{eqnarray}
\Sigma r^2\Omega(\dot\gamma+\bar
v_r\gamma^\prime)\sin\beta&=&Y\nonumber\\ &=&Q_1\sI
r\Omega^2\gamma^\prime\sin\beta+{{1}\over{r}}\partial_r\left[\sI
r^3\Omega^2(Q_2\gamma^\prime\sin\beta+Q_3\beta^\prime)\right]+\sI
r^2\Omega^2\gamma^\prime\cos\beta(Q_2\beta^\prime-
Q_3\gamma^\prime\sin\beta).\label{new3}
\end{eqnarray}
It will be useful to consider the complex combination
\begin{equation}
X+{\rm i}Y=Q_1\sI\Omega^2\psi+{{1}\over{r}}(\partial_r+{\rm
i}\gamma^\prime\cos\beta)(Q_4\sI r^2\Omega^2\psi),
\end{equation}
where $Q_4=Q_2+{\rm i}Q_3$.

\section{Non-linear bending waves in a thin disc}

\subsection{Asymptotic expansions}

Consider a thin disc in a spherically symmetric potential $\Phi(r)$,
subject to a slowly varying, unforced warp of large amplitude.  Let
the small parameter $\epsilon$ be a characteristic value of the local
angular semi-thickness of the disc.  Then, to resolve the internal
structure of the disc, introduce the stretched vertical coordinate
$\zeta$ by
\begin{equation}
\theta={{\pi}\over{2}}-\epsilon\zeta,
\end{equation}
so that $\zeta=O(1)$ within the disc.  All quantities vary not on the
fast, orbital time-scale, but on the slow time-scale characteristic of
both viscous and dynamical evolution of the warp.  This is captured by
the slow time coordinate
\begin{equation}
T=\epsilon^2t.
\end{equation}
[It is implicit that units are chosen such that the radius of the disc
and the orbital time-scale are $O(1)$.]  Then set
\begin{eqnarray}
\beta(r,t)&\mapsto&\beta(r,T),\\
\gamma(r,t)&\mapsto&\gamma(r,T).
\end{eqnarray}
A dot will now denote differentiation with respect to $T$.  For the
density and pressure, introduce the scalings
\begin{eqnarray}
\rho(r,\theta,\phi,t)&=&\epsilon^s\left[\rho_0(r,\phi,\zeta,T)+
\epsilon\rho_1(r,\phi,\zeta,T)+O(\epsilon^2)\right],\\
p(r,\theta,\phi,t)&=&\epsilon^{s+2}\left[p_0(r,\phi,\zeta,T)+\epsilon
p_1(r,\phi,\zeta,T)+O(\epsilon^2)\right],
\end{eqnarray}
where $s$ is a parameter which should be positive if the
self-gravitation of the disc is to be negligible, but otherwise has no
effect on the following analysis (cf.~Ogilvie 1997).  For the
velocities, set
\begin{eqnarray}
v_r(r,\theta,\phi,t)&=&\epsilon
v_{r1}(r,\phi,\zeta,T)+\epsilon^2v_{r2}(r,\phi,\zeta,T)+O(\epsilon^3),\\
v_\theta(r,\theta,\phi,t)&=&\epsilon
v_{\theta1}(r,\phi,\zeta,T)+\epsilon^2v_{\theta2}(r,\phi,\zeta,T)+
O(\epsilon^3),\\
v_\phi(r,\theta,\phi,t)&=&r\Omega(r)\sin\theta+\epsilon
v_{\phi1}(r,\phi,\zeta,T)+\epsilon^2
v_{\phi2}(r,\phi,\zeta,T)+O(\epsilon^3).
\end{eqnarray}
Finally, for the viscosities, assume
\begin{eqnarray}
\mu(r,\theta,\phi,t)&=&\epsilon^{s+2}\left[\mu_0(r,\phi,\zeta,T)+
\epsilon\mu_1(r,\phi,\zeta,T)+O(\epsilon^2)\right],\\
\mu_{\rm b}(r,\theta,\phi,t)&=&\epsilon^{s+2}\left[\mu_{{\rm
b}0}(r,\phi,\zeta,T)+\epsilon\mu_{{\rm
b}1}(r,\phi,\zeta,T)+O(\epsilon^2)\right].
\end{eqnarray}
This is the correct scaling for an $\alpha$-viscosity (Shakura \&
Sunyaev 1973) with $\alpha=O(1)$, which includes the possibility of
small or zero $\alpha$ unless resonance occurs.  Note that, for this
non-linear warp, all quantities other than the orbital angular
velocity (and the surface density; see below) are non-axisymmetric at
leading order in $\epsilon$.

These expansions may then be substituted into the dynamical equations.
First, equation (\ref{dur}) at $O(\epsilon^s)$ yields
\begin{equation}
-\rho_0r\Omega^2=-\rho_0\Phi^\prime,
\end{equation}
which determines the orbital angular velocity.  The epicyclic
frequency $\kappa(r)$ is defined by
\begin{equation}
\kappa^2=4\Omega^2+2r\Omega\Omega^\prime,
\end{equation}
and the dimensionless epicyclic frequency is
$\tilde\kappa=\kappa/\Omega$.  The remaining equations may be divided
into two sets.

\subsection{Set~A equations}

Equation (\ref{drho}) at $O(\epsilon^s)$:
\begin{equation}
\left(\Omega\partial_\phi-{{v_{\theta1}}\over{r}}\partial_\zeta\right)
\rho_0={{\rho_0}\over{r}}\partial_\zeta
v_{\theta1}.\label{a1}
\end{equation}
Equation (\ref{dp}) at $O(\epsilon^{s+2})$:
\begin{equation}
\left(\Omega\partial_\phi-{{v_{\theta1}}\over{r}}\partial_\zeta\right)
p_0={{\Gamma
p_0}\over{r}}\partial_\zeta v_{\theta1}.\label{a2}
\end{equation}
Equation (\ref{dur}) at $O(\epsilon^{s+1})$:
\begin{eqnarray}
\lefteqn{\rho_0\left(\Omega\partial_\phi-{{v_{\theta1}}\over{r}}
\partial_\zeta\right)v_{r1}-2\rho_0\Omega(v_{\phi1}+rv_{r1}
\gamma^\prime\cos\beta)=-(\beta^\prime\cos\phi+\gamma^\prime
\sin\beta\sin\phi)\partial_\zeta\left[p_0+(\mu_{{\rm
b}0}+{\textstyle{{1}\over{3}}}\mu_0){{1}\over{r}}\partial_\zeta
v_{\theta1}\right]}&\nonumber\\
&&+\left[{{1}\over{r^2}}+(\beta^\prime\cos\phi+\gamma^\prime
\sin\beta\sin\phi)^2\right]\partial_\zeta(\mu_0\partial_\zeta
v_{r1})+\Omega(\beta^\prime\sin\phi-\gamma^\prime\sin\beta\cos\phi)
\partial_\zeta\mu_0.\label{a3}
\end{eqnarray}
Equation (\ref{dutheta}) at $O(\epsilon^{s+1})$:
\begin{eqnarray}
\lefteqn{\rho_0\left(\Omega\partial_\phi-{{v_{\theta1}}\over{r}}
\partial_\zeta\right)\left[v_{\theta1}+rv_{r1}(\beta^\prime\cos\phi+
\gamma^\prime\sin\beta\sin\phi)\right]}&\nonumber\\
&&-\rho_0\Omega\left[r\Omega\zeta+rv_{r1}(\beta^\prime\sin\phi-
\gamma^\prime\sin\beta\cos\phi)\right]={{1}\over{r}}\partial_\zeta
\left[p_0+(\mu_{{\rm
b}0}+{\textstyle{{1}\over{3}}}\mu_0){{1}\over{r}}\partial_\zeta
v_{\theta1}\right]\nonumber\\
&&+\left[{{1}\over{r^2}}+(\beta^\prime\cos\phi+\gamma^\prime
\sin\beta\sin\phi)^2\right]\partial_\zeta\left\{\mu_0
\partial_\zeta\left[v_{\theta1}+rv_{r1}(\beta^\prime\cos\phi+
\gamma^\prime\sin\beta\sin\phi)\right]\right\}\nonumber\\
&&-r\Omega(\beta^\prime\cos\phi+\gamma^\prime\sin\beta\sin\phi)
(\beta^\prime\sin\phi-\gamma^\prime\sin\beta\cos\phi)
\partial_\zeta\mu_0.\label{a4}
\end{eqnarray}
Equation (\ref{duphi}) at $O(\epsilon^{s+1})$:
\begin{eqnarray}
\lefteqn{\rho_0\left(\Omega\partial_\phi-{{v_{\theta1}}\over{r}}
\partial_\zeta\right)(v_{\phi1}+rv_{r1}\gamma^\prime\cos\beta)+
{{\rho_0\kappa^2}\over{2\Omega}}v_{r1}=\left[{{1}\over{r^2}}+
(\beta^\prime\cos\phi+\gamma^\prime\sin\beta\sin\phi)^2\right]
\partial_\zeta\left[\mu_0\partial_\zeta(v_{\phi1}+rv_{r1}
\gamma^\prime\cos\beta)\right]}\nonumber\\
&&+r\Omega^\prime(\beta^\prime\cos\phi+\gamma^\prime\sin\beta\sin\phi)
\partial_\zeta\mu_0.\label{a5}
\end{eqnarray}

\subsection{Set~B equations}

Equation (\ref{drho}) at $O(\epsilon^{s+1})$:
\begin{equation}
\left(\Omega\partial_\phi-{{v_{\theta1}}\over{r}}\partial_\zeta\right)
\rho_1+\left(v_{r1}\partial_r-{{v_{\theta2}}\over{r}}\partial_\zeta+
{{v_{\phi1}}\over{r}}\partial_\phi\right)\rho_0={{\rho_1}\over{r}}
\partial_\zeta
v_{\theta1}-\rho_0\left[{{1}\over{r^2}}\partial_r(r^2v_{r1})-
{{1}\over{r}}\partial_\zeta
v_{\theta2}+{{1}\over{r}}\partial_\phi v_{\phi1}\right].\label{b1}
\end{equation}
Equation (\ref{dp}) at $O(\epsilon^{s+3})$:
\begin{equation}
\left(\Omega\partial_\phi-{{v_{\theta1}}\over{r}}\partial_\zeta\right)
p_1+\left(v_{r1}\partial_r-{{v_{\theta2}}\over{r}}\partial_\zeta+
{{v_{\phi1}}\over{r}}\partial_\phi\right)p_0={{\Gamma p_1}\over{r}}
\partial_\zeta v_{\theta1}-\Gamma p_0\left[{{1}\over{r^2}}
\partial_r(r^2v_{r1})-{{1}\over{r}}\partial_\zeta v_{\theta2}+
{{1}\over{r}}\partial_\phi v_{\phi1}\right].\label{b2}
\end{equation}
Equation (\ref{dur}) at $O(\epsilon^{s+2})$:
\begin{eqnarray}
\lefteqn{\rho_0\left(\Omega\partial_\phi-{{v_{\theta1}}\over{r}}
\partial_\zeta\right)v_{r2}+\rho_0\left(v_{r1}\partial_r-
{{v_{\theta2}}\over{r}}\partial_\zeta+{{v_{\phi1}}\over{r}}\partial_\phi
\right)v_{r1}+\rho_1\left(\Omega\partial_\phi-{{v_{\theta1}}\over{r}}
\partial_\zeta\right)v_{r1}-{{\rho_0}\over{r}}\left[v_{\theta1}+
rv_{r1}(\beta^\prime\cos\phi+\gamma^\prime\sin\beta\sin\phi)\right]^2}
&\nonumber\\
&&-2\rho_0\Omega\left[-{\textstyle{{1}\over{2}}}r\Omega\zeta^2+
v_{\phi2}+r(\dot\gamma+v_{r2}\gamma^\prime)\cos\beta-rv_{r1}
(\beta^\prime\sin\phi-\gamma^\prime\sin\beta\cos\phi)\zeta\right]-
{{\rho_0}\over{r}}(v_{\phi1}+rv_{r1}\gamma^\prime\cos\beta)^2
\nonumber\\ &&-2\rho_1\Omega(v_{\phi1}+rv_{r1}\gamma^\prime\cos\beta)=
-(\beta^\prime\cos\phi+\gamma^\prime\sin\beta\sin\phi)\partial_\zeta
\left\{p_1-(\mu_{{\rm
b}0}+{\textstyle{{1}\over{3}}}\mu_0)\left[{{1}\over{r^2}}
\partial_r(r^2v_{r1})-{{1}\over{r}}\partial_\zeta
v_{\theta2}+{{1}\over{r}}\partial_\phi
v_{\phi1}\right]\right.\nonumber\\ &&\left.+(\mu_{{\rm
b}1}+{\textstyle{{1}\over{3}}}\mu_1){{1}\over{r}}\partial_\zeta
v_{\theta1}\right\}-(\partial_r-\gamma^\prime\cos\beta\,\partial_\phi)
\left[p_0+(\mu_{{\rm
b}0}+{\textstyle{{1}\over{3}}}\mu_0){{1}\over{r}}\partial_\zeta
v_{\theta1}\right]\nonumber\\
&&+\left[{{1}\over{r^2}}+(\beta^\prime\cos\phi+\gamma^\prime
\sin\beta\sin\phi)^2\right]\partial_\zeta(\mu_0\partial_\zeta
v_{r2}+\mu_1\partial_\zeta
v_{r1})+(\beta^\prime\cos\phi+\gamma^\prime\sin\beta\sin\phi)
\partial_\zeta\left[\mu_0(\partial_r-\gamma^\prime\cos\beta\,
\partial_\phi)v_{r1}\right]\nonumber\\
&&+{{1}\over{r^2}}(\partial_r-\gamma^\prime\cos\beta\,\partial_\phi)
\left[\mu_0r^2(\beta^\prime\cos\phi+\gamma^\prime\sin\beta\sin\phi)
\partial_\zeta
v_{r1}\right]-{{2v_{r1}}\over{r}}(\beta^\prime\cos\phi+\gamma^\prime
\sin\beta\sin\phi)\partial_\zeta\mu_0\nonumber\\
&&+{{1}\over{r^3}}\left[(\partial_r-\gamma^\prime\cos\beta\,\partial_\phi)
(\mu_0r^2)\right]\partial_\zeta\left[v_{\theta1}+rv_{r1}
(\beta^\prime\cos\phi+\gamma^\prime\sin\beta\sin\phi)\right]\nonumber\\
&&-(\partial_\zeta\mu_0)(\partial_r-\gamma^\prime\cos\beta\,
\partial_\phi)\left[{{v_{\theta1}}\over{r}}+v_{r1}(\beta^\prime\cos\phi+
\gamma^\prime\sin\beta\sin\phi)\right]\nonumber\\
&&-{{1}\over{r}}(\partial_\zeta\mu_0)\partial_\phi
\left[(\beta^\prime\cos\phi+\gamma^\prime\sin\beta\sin\phi)(v_{\phi1}+
rv_{r1}\gamma^\prime\cos\beta)\right]+\Omega^\prime\partial_\phi\mu_0
\nonumber\\
&&+{{1}\over{r}}(\beta^\prime\cos\phi+\gamma^\prime\sin\beta\sin\phi)
(\partial_\phi\mu_0)\partial_\zeta(v_{\phi1}+rv_{r1}
\gamma^\prime\cos\beta)+\Omega(\beta^\prime\sin\phi-\gamma^\prime
\sin\beta\cos\phi)\partial_\zeta\mu_1.\label{b3}
\end{eqnarray}
Equation (\ref{dutheta}) at $O(\epsilon^{s+2})$:
\begin{eqnarray}
\lefteqn{\rho_0\left(\Omega\partial_\phi-{{v_{\theta1}}\over{r}}
\partial_\zeta\right)\left[v_{\theta2}+r(\dot\beta+v_{r2}\beta^\prime)
\cos\phi+r(\dot\gamma+v_{r2}\gamma^\prime)\sin\beta\sin\phi\right]}
&\nonumber\\
&&+\rho_0\left(v_{r1}\partial_r-{{v_{\theta2}}\over{r}}\partial_\zeta+
{{v_{\phi1}}\over{r}}\partial_\phi\right)\left[v_{\theta1}+rv_{r1}
(\beta^\prime\cos\phi+\gamma^\prime\sin\beta\sin\phi)\right]\nonumber\\
&&+\rho_1\left(\Omega\partial_\phi-{{v_{\theta1}}\over{r}}
\partial_\zeta\right)\left[v_{\theta1}+rv_{r1}(\beta^\prime\cos\phi+
\gamma^\prime\sin\beta\sin\phi)\right]\nonumber+{{\rho_0v_{r1}}\over{r}}
\left[v_{\theta1}+rv_{r1}(\beta^\prime\cos\phi+\gamma^\prime
\sin\beta\sin\phi)\right]\nonumber\\
&&-\rho_0\Omega\left[(v_{\phi1}+rv_{r1}\gamma^\prime\cos\beta)\zeta+
r(\dot\beta+v_{r2}\beta^\prime)\sin\phi-r(\dot\gamma+v_{r2}\gamma^\prime)
\sin\beta\cos\phi\right]\nonumber\\
&&-{{\rho_0}\over{r}}(v_{\phi1}+rv_{r1}\gamma^\prime\cos\beta)
\left[r\Omega\zeta+rv_{r1}(\beta^\prime\sin\phi-\gamma^\prime
\sin\beta\cos\phi)\right]-\rho_1\Omega\left[r\Omega\zeta+rv_{r1}
(\beta^\prime\sin\phi-\gamma^\prime\sin\beta\cos\phi)\right]\nonumber\\
&&={{1}\over{r}}\partial_\zeta\left\{p_1-(\mu_{{\rm
b}0}+{\textstyle{{1}\over{3}}}\mu_0)\left[{{1}\over{r^2}}
\partial_r(r^2v_{r1})-{{1}\over{r}}\partial_\zeta
v_{\theta2}+{{1}\over{r}}\partial_\phi v_{\phi1}\right]+(\mu_{{\rm
b}1}+{\textstyle{{1}\over{3}}}\mu_1){{1}\over{r}}\partial_\zeta
v_{\theta1}\right\}\nonumber\\
&&+\left[{{1}\over{r^2}}+(\beta^\prime\cos\phi+\gamma^\prime\sin\beta
\sin\phi)^2\right]\partial_\zeta\left\{\mu_0\partial_\zeta
\left[v_{\theta2}+r(\dot\beta+v_{r2}\beta^\prime)\cos\phi+
r(\dot\gamma+v_{r2}\gamma^\prime)\sin\beta\sin\phi\right]\right.\nonumber\\
&&\left.+\mu_1\partial_\zeta\left[v_{\theta1}+rv_{r1}
(\beta^\prime\cos\phi+\gamma^\prime\sin\beta\sin\phi)\right]\right\}
\nonumber\\
&&+(\beta^\prime\cos\phi+\gamma^\prime\sin\beta\sin\phi)\partial_\zeta
\left\{\mu_0(\partial_r-\gamma^\prime\cos\beta\,\partial_\phi)
\left[v_{\theta1}+rv_{r1}(\beta^\prime\cos\phi+\gamma^\prime
\sin\beta\sin\phi)\right]\right\}\nonumber\\
&&+{{1}\over{r^2}}(\partial_r-\gamma^\prime\cos\beta\,\partial_\phi)
\left\{\mu_0r^2(\beta^\prime\cos\phi+\gamma^\prime\sin\beta\sin\phi)
\partial_\zeta\left[v_{\theta1}+rv_{r1}(\beta^\prime\cos\phi+
\gamma^\prime\sin\beta\sin\phi)\right]\right\}\nonumber\\
&&-{{1}\over{r}}(\beta^\prime\cos\phi+\gamma^\prime\sin\beta\sin\phi)
(\partial_\zeta\mu_0)\left[v_{\theta1}+rv_{r1}(\beta^\prime\cos\phi+
\gamma^\prime\sin\beta\sin\phi)\right]-{{1}\over{r^3}}(\partial_\zeta
v_{r1})(\partial_r-\gamma^\prime\cos\beta\,\partial_\phi)(\mu_0r^2)
\nonumber\\
&&+{{1}\over{r}}(\partial_\zeta\mu_0)(\partial_r-\gamma^\prime\cos\beta\,
\partial_\phi)v_{r1}+{{1}\over{r^2}}(\partial_\zeta\mu_0)
\partial_\phi(v_{\phi1}+rv_{r1}\gamma^\prime\cos\beta)\nonumber\\
&&-{{1}\over{r^2}}(\partial_\phi\mu_0)\partial_\zeta(v_{\phi1}+
rv_{r1}\gamma^\prime\cos\beta)-(\beta^\prime\sin\phi-\gamma^\prime
\sin\beta\cos\phi)(\beta^\prime\cos\phi+\gamma^\prime\sin\beta\sin\phi)
\partial_\zeta\left[\mu_0(v_{\phi1}+rv_{r1}\gamma^\prime\cos\beta)\right]
\nonumber\\
&&-r\Omega(\beta^\prime\sin\phi-\gamma^\prime\sin\beta\cos\phi)
(\beta^\prime\cos\phi+\gamma^\prime\sin\beta\sin\phi)\partial_\zeta\mu_1-
{{1}\over{r^2}}(\partial_r-\gamma^\prime\cos\beta\,\partial_\phi)
\left[\mu_0r^3\Omega(\beta^\prime\sin\phi-\gamma^\prime\sin\beta\cos\phi)
\right]\nonumber\\
&&-\mu_0(\beta^\prime\sin\phi-\gamma^\prime\sin\beta\cos\phi)
\left[(r\Omega)^\prime+(\beta^\prime\cos\phi+\gamma^\prime\sin\beta
\sin\phi)\partial_\zeta(v_{\phi1}+rv_{r1}\gamma^\prime\cos\beta)\right].
\label{b4}
\end{eqnarray}
Equation (\ref{duphi}) at $O(\epsilon^{s+2})$:
\begin{eqnarray}
\lefteqn{\rho_0\left(\Omega\partial_\phi-{{v_{\theta1}}\over{r}}
\partial_\zeta\right)\left[v_{\phi2}-{\textstyle{{1}\over{2}}}
r\Omega\zeta^2+r(\dot\gamma+v_{r2}\gamma^\prime\cos\beta)-
rv_{r1}(\beta^\prime\sin\phi-\gamma^\prime\sin\beta\cos\phi)
\zeta\right]}&\nonumber\\
&&+\rho_0\left(v_{r1}\partial_r-{{v_{\theta2}}\over{r}}\partial_\zeta+
{{v_{\phi1}}\over{r}}\partial_\phi\right)(v_{\phi1}+rv_{r1}
\gamma^\prime\cos\beta)+\rho_1\left(\Omega\partial_\phi-
{{v_{\theta1}}\over{r}}\partial_\zeta\right)(v_{\phi1}+rv_{r1}
\gamma^\prime\cos\beta)+{{\rho_0\kappa^2}\over{2\Omega}}v_{r2}+
{{\rho_1\kappa^2}\over{2\Omega}}v_{r1}\nonumber\\
&&+{{\rho_0}\over{r}}v_{r1}(v_{\phi1}+rv_{r1}\gamma^\prime\cos\beta)+
{{\rho_0}\over{r}}\left[v_{\theta1}+rv_{r1}(\beta^\prime\cos\phi+
\gamma^\prime\sin\beta\sin\phi)\right]\left[r\Omega\zeta+rv_{r1}
(\beta^\prime\sin\phi-\gamma^\prime\sin\beta\cos\phi)\right]\nonumber\\
&&=-{{1}\over{r}}\partial_\phi\left[p_0+(\mu_{{\rm
b}0}+{\textstyle{{1}\over{3}}}\mu_0){{1}\over{r}}\partial_\zeta
v_{\theta1}\right]+r\Omega^\prime(\beta^\prime\cos\phi+\gamma^\prime
\sin\beta\sin\phi)\partial_\zeta\mu_1\nonumber\\
&&+{{1}\over{r^2}}\partial_r\left[\mu_0r^2(r\Omega)^\prime\right]-
r\Omega^\prime(\partial_\phi\mu_0)\gamma^\prime\cos\beta+
(\beta^\prime\cos\phi+\gamma^\prime\sin\beta\sin\phi)\partial_\zeta
\left[\mu_0(\partial_r-\gamma^\prime\cos\beta\,\partial_\phi)
(v_{\phi1}+rv_{r1}\gamma^\prime\cos\beta)\right]\nonumber\\
&&+{{1}\over{r^2}}(\partial_r-\gamma^\prime\cos\beta\,\partial_\phi)
\left[\mu_0r^2(\beta^\prime\cos\phi+\gamma^\prime\sin\beta\sin\phi)
\partial_\zeta(v_{\phi1}+rv_{r1}\gamma^\prime\cos\beta)\right]\nonumber\\
&&+\left[{{1}\over{r^2}}+(\beta^\prime\cos\phi+\gamma^\prime
\sin\beta\sin\phi)^2\right]\partial_\zeta\left\{\mu_0\partial_\zeta
\left[-{\textstyle{{1}\over{2}}}r\Omega\zeta^2+v_{\phi2}+
r(\dot\gamma+v_{r2}\gamma^\prime)\cos\beta-rv_{r1}(\beta^\prime\sin\phi-
\gamma^\prime\sin\beta\cos\phi)\zeta\right]\right.\nonumber\\
&&\left.+\mu_1\partial_\zeta(v_{\phi1}+rv_{r1}\gamma^\prime\cos\beta)
\right\}-{{\Omega}\over{r}}\partial_r(\mu_0r)-{{1}\over{r}}
(v_{\phi1}+rv_{r1}\gamma^\prime\cos\beta)(\beta^\prime\cos\phi+
\gamma^\prime\sin\beta\sin\phi)\partial_\zeta\mu_0\nonumber\\
&&+{{\Omega}\over{r}}\zeta\partial_\zeta\mu_0+{{1}\over{r}}
(\partial_\phi
v_{r1})(\beta^\prime\cos\phi+\gamma^\prime\sin\beta\sin\phi)
\partial_\zeta\mu_0-{{1}\over{r}}(\partial_\phi\mu_0)
(\beta^\prime\cos\phi+\gamma^\prime\sin\beta\sin\phi)\partial_\zeta
v_{r1}\nonumber\\
&&-{{1}\over{r^2}}(\partial_\zeta\mu_0)\partial_\phi\left[v_{\theta1}+
rv_{r1}(\beta^\prime\cos\phi+\gamma^\prime\sin\beta\sin\phi)\right]+
{{1}\over{r^2}}(\partial_\phi\mu_0)\partial_\zeta\left[v_{\theta1}+
rv_{r1}(\beta^\prime\cos\phi+\gamma^\prime\sin\beta\sin\phi)\right]
\nonumber\\
&&+(\beta^\prime\sin\phi-\gamma^\prime\sin\beta\cos\phi)
(\beta^\prime\cos\phi+\gamma^\prime\sin\beta\sin\phi)\partial_\zeta
\left\{\mu_0\left[v_{\theta1}+rv_{r1}(\beta^\prime\cos\phi+
\gamma^\prime\sin\beta\sin\phi)\right]\right\}\nonumber\\
&&+\mu_0(\beta^\prime\sin\phi-\gamma^\prime\sin\beta\cos\phi)
(\beta^\prime\cos\phi+\gamma^\prime\sin\beta\sin\phi)\partial_\zeta
\left[v_{\theta1}+rv_{r1}(\beta^\prime\cos\phi+\gamma^\prime
\sin\beta\sin\phi)\right]\nonumber\\
&&-\mu_0r\Omega(\beta^\prime\sin\phi-\gamma^\prime\sin\beta\cos\phi)^2.
\label{b5}
\end{eqnarray}

\subsection{Integrated quantities}

Finally, equation (\ref{drho}) is also required at
$O(\epsilon^{s+2})$, but only in its integrated
form,\footnote{Throughout this paper, integrations with respect to
$\phi$ are carried out from $0$ to $2\pi$, and integrations with
respect to $\zeta$ are carried out over the full vertical extent of
the disc.}
\begin{equation}
\partial_T\left(\int\!\!\!\int\rho_0\,r\,{\rm d}\phi\,{\rm
d}\zeta\right)+{{1}\over{r}}\partial_r\left[r\int\!\!\!\int(\rho_0v_{r2}+
\rho_1v_{r1})\,r\,{\rm
d}\phi\,{\rm d}\zeta\right]=0.\label{sigma}
\end{equation}

The surface density $\Sigma(r,T)$ at leading order in $\epsilon$ is
\begin{equation}
\Sigma=\int\rho_0\,r\,{\rm d}\zeta,
\end{equation}
and is independent of $\phi$ by virtue of equation (\ref{a1}).  Other
vertically integrated quantities are non-axisymmetric, however, and
these will be written with tildes.  The corresponding azimuthally
averaged quantities, written without tildes, are defined by the
operation
\begin{equation}
\langle\,\cdot\,\rangle={{1}\over{2\pi}}\int\,\cdot\,\,{\rm d}\phi.
\end{equation}
Thus a suitably averaged radial velocity $\bar v_r(r,T)$ may be defined by
\begin{equation}
\Sigma\bar v_r=\sF=\langle\tilde{\sF}\rangle,
\end{equation}
where
\begin{equation}
\tilde{\sF}=\int(\rho_0v_{r2}+\rho_1v_{r1})\,r\,{\rm d}\zeta
\end{equation}
is the radial mass flux at $O(\epsilon^{s+3})$.\footnote{There is a
radial mass flux at $O(\epsilon^{s+2})$, but its azimuthal average
vanishes by virtue of equations (\ref{a5}) and (\ref{a1}).}  Then
equation (\ref{sigma}), divided by $2\pi$, reads
\begin{equation}
\partial_T\Sigma+{{1}\over{r}}\partial_r(r\Sigma\bar v_r)=0,
\end{equation}
which agrees with equation (\ref{jep1}).  Two more definitions will be
useful.  A suitably averaged kinematic viscosity $\bar\nu(r,T)$ may be
defined by
\begin{equation}
\bar\nu\Sigma=\sV=\langle\tilde{\sV}\rangle
\end{equation}
where
\begin{equation}
\tilde{\sV}=\int\mu_0\,r\,{\rm d}\zeta
\end{equation}
is the vertically integrated viscosity.  Finally, the second vertical
moment of the density is
\begin{equation}
\tilde{\sI}=\int\rho_0r^2\zeta^2\,r\,{\rm d}\zeta,
\end{equation}
and its azimuthal average is
\begin{equation}
\sI=\langle\tilde{\sI}\rangle.
\end{equation}

\subsection{Formal manipulations}

It is clear that the problem under consideration is much more
complicated than the analysis of an unwarped viscous disc.  In the
absence of a warp, all quantities are independent of $\phi$ and
symmetric about $\zeta=0$.  Furthermore, the quantities $v_{r1}$,
$v_{\theta1}$, $v_{\phi1}$ and $v_{\theta2}$ all vanish.  It is worth
noting that {\it if the same assumptions could be made for a warped
disc\/}, equations (\ref{jep3}) and (\ref{jep4}) with
$\nu_1=\nu_2=\bar\nu$ could be derived directly by integrating
equations (\ref{b5}) and (\ref{b4}).  However, these assumptions are
inconsistent with the equations of Set~A.  For example, in equation
(\ref{a3}) there is a radial force resulting from the lateral
imbalance of pressure and viscous stress in a warped disc, and this
drives the horizontal velocities $v_{r1}$ and $v_{\phi1}$.

It is important to note the formal structure of the problem.  Set~A
consists of five coupled non-linear partial differential equations
(PDEs) with two independent variables $\{\phi,\zeta\}$ and seven
dependent variables $\{\rho_0,p_0,\mu_0,\mu_{{\rm
b}0},v_{r1},v_{\theta1},v_{\phi1}\}$.  The equations must be closed by
specifying suitable prescriptions for the viscosities and for the
thermodynamics.  In general, they must be solved numerically.  Set~B
is a set of five coupled linear PDEs for the higher-order quantities
$\{\rho_1,p_1,\mu_1,\mu_{{\rm b}1},v_{r2},v_{\theta2},v_{\phi2}\}$,
with coefficients that depend on the solutions of Set~A and their
radial derivatives.  Fortunately, all the information required from
Set~B can be extracted by integration, so that only Set~A need be
solved in detail.

The aim is to extract equations resembling equations
(\ref{new1})--(\ref{new3}).  To this end, multiply equation (\ref{b5})
by $r$, integrate vertically and average azimuthally to obtain
\begin{equation}
\Sigma\bar v_r(r^2\Omega)^\prime=I_1,
\end{equation}
after an integration by parts and the use of equations (\ref{a1}) and
(\ref{b1}), where
\begin{eqnarray}
\lefteqn{I_1=\int\big\langle{{1}\over{r^2}}\partial_r\left
[\mu_0r^4\Omega^\prime-\rho_0r^3v_{r1}(v_{\phi1}+rv_{r1}\gamma^\prime
\cos\beta)+\mu_0r^3(\beta^\prime\cos\phi+\gamma^\prime\sin\beta\sin\phi)
\partial_\zeta(v_{\phi1}+rv_{r1}\gamma^\prime\cos\beta)\right]}&\nonumber\\
&&-\rho_0\left[v_{\theta1}+rv_{r1}(\beta^\prime\cos\phi+\gamma^\prime
\sin\beta\sin\phi)\right]\left[r\Omega\zeta+rv_{r1}(\beta^\prime\sin\phi-
\gamma^\prime\sin\beta\cos\phi)\right]-\mu_0(\beta^\prime\sin\phi-
\gamma^\prime\sin\beta\cos\phi)\partial_\zeta
v_{r1}\nonumber\\
&&+\mu_0r(\beta^\prime\cos\phi+\gamma^\prime\sin\beta\sin\phi)
(\beta^\prime\sin\phi-\gamma^\prime\sin\beta\cos\phi)\partial_\zeta
\left[v_{\theta1}+rv_{r1}(\beta^\prime\cos\phi+\gamma^\prime
\sin\beta\sin\phi)\right]\nonumber\\
&&-\mu_0r^2\Omega(\beta^\prime\sin\phi-\gamma^\prime\sin\beta\cos\phi)^2
\big\rangle\,r\,{\rm
d}\zeta.\label{I1}
\end{eqnarray}
Then multiply equation (\ref{b4}) first by $(-r\sin\phi)$, then by
$r\cos\phi$, in each case integrating vertically and averaging
azimuthally to obtain first
\begin{equation}
\Sigma r^2\Omega(\dot\beta+\bar v_r\beta^\prime)=I_2,
\end{equation}
and then
\begin{equation}
\Sigma r^2\Omega(\dot\gamma+\bar v_r\gamma^\prime)\sin\beta=I_3,
\end{equation}
where
\begin{eqnarray}
\lefteqn{I_2=\int\big\langle
r^2\Omega\zeta\cos\phi\left[{{1}\over{r^2}}\partial_r(r^2\rho_0v_{r1})+
{{1}\over{r}}\partial_\phi(\rho_0v_{\phi1})\right]+{{\sin\phi}\over{r^2}}
\partial_r\left\{r^3\rho_0v_{r1}\left[v_{\theta1}+rv_{r1}
(\beta^\prime\cos\phi+\gamma^\prime\sin\beta\sin\phi)\right]\right\}}&
\nonumber\\
&&-\rho_0v_{\phi1}\cos\phi\left[v_{\theta1}+rv_{r1}(\beta^\prime\cos\phi+
\gamma^\prime\sin\beta\sin\phi)\right]-2\rho_0r\Omega\zeta\sin\phi
(v_{\phi1}+rv_{r1}\gamma^\prime\cos\beta)\nonumber\\
&&-\rho_0rv_{r1}\sin\phi(v_{\phi1}+rv_{r1}\gamma^\prime\cos\beta)
(\beta^\prime\sin\phi-\gamma^\prime\sin\beta\cos\phi)\nonumber\\
&&-{{\sin\phi}\over{r}}(\partial_r-\gamma^\prime\cos\beta\,\partial_\phi)
\left\{\mu_0r^2(\beta^\prime\cos\phi+\gamma^\prime\sin\beta\sin\phi)
\partial_\zeta\left[v_{\theta1}+rv_{r1}(\beta^\prime\cos\phi+
\gamma^\prime\sin\beta\sin\phi)\right]\right\}\nonumber\\
&&-\mu_0\sin\phi(\beta^\prime\cos\phi+\gamma^\prime\sin\beta\sin\phi)
\partial_\zeta\left[v_{\theta1}+rv_{r1}(\beta^\prime\cos\phi+
\gamma^\prime\sin\beta\sin\phi)\right]+{{\sin\phi}\over{r^2}}
\partial_r(\mu_0r^2\partial_\zeta
v_{r1})-{{\mu_0\cos\phi}\over{r}}\partial_\zeta v_{\phi1}\nonumber\\
&&+{{\sin\phi}\over{r}}\partial_r\left[\mu_0r^3\Omega
(\beta^\prime\sin\phi-\gamma^\prime\sin\beta\cos\phi)\right]+
\mu_0r^2\Omega\gamma^\prime\cos\beta\cos\phi(\beta^\prime\sin\phi-
\gamma^\prime\sin\beta\cos\phi)\nonumber\\
&&+\mu_0r\sin\phi(\beta^\prime\sin\phi-\gamma^\prime\sin\beta\cos\phi)
\left[(r\Omega)^\prime+(\beta^\prime\cos\phi+\gamma^\prime
\sin\beta\sin\phi)\partial_\zeta(v_{\phi1}+rv_{r1}\gamma^\prime
\cos\beta)\right]\big\rangle\,r\,{\rm
d}\zeta
\end{eqnarray}
and
\begin{eqnarray}
\lefteqn{I_3=\int\big\langle
r^2\Omega\zeta\sin\phi\left[{{1}\over{r^2}}\partial_r(r^2\rho_0v_{r1})+
{{1}\over{r}}\partial_\phi(\rho_0v_{\phi1})\right]-{{\cos\phi}\over{r^2}}
\partial_r\left\{r^3\rho_0v_{r1}\left[v_{\theta1}+rv_{r1}(\beta^\prime
\cos\phi+\gamma^\prime\sin\beta\sin\phi)\right]\right\}}&\nonumber\\
&&-\rho_0v_{\phi1}\sin\phi\left[v_{\theta1}+rv_{r1}(\beta^\prime\cos\phi+
\gamma^\prime\sin\beta\sin\phi)\right]+2\rho_0r\Omega\zeta\cos\phi
(v_{\phi1}+rv_{r1}\gamma^\prime\cos\beta)\nonumber\\
&&+\rho_0rv_{r1}\cos\phi(v_{\phi1}+rv_{r1}\gamma^\prime\cos\beta)
(\beta^\prime\sin\phi-\gamma^\prime\sin\beta\cos\phi)\nonumber\\
&&+{{\cos\phi}\over{r}}(\partial_r-\gamma^\prime\cos\beta\,\partial_\phi)
\left\{\mu_0r^2(\beta^\prime\cos\phi+\gamma^\prime\sin\beta\sin\phi)
\partial_\zeta\left[v_{\theta1}+rv_{r1}(\beta^\prime\cos\phi+
\gamma^\prime\sin\beta\sin\phi)\right]\right\}\nonumber\\
&&+\mu_0\cos\phi(\beta^\prime\cos\phi+\gamma^\prime\sin\beta\sin\phi)
\partial_\zeta\left[v_{\theta1}+rv_{r1}(\beta^\prime\cos\phi+
\gamma^\prime\sin\beta\sin\phi)\right]-{{\cos\phi}\over{r^2}}
\partial_r(\mu_0r^2\partial_\zeta
v_{r1})-{{\mu_0\sin\phi}\over{r}}\partial_\zeta v_{\phi1}\nonumber\\
&&-{{\cos\phi}\over{r}}\partial_r\left[\mu_0r^3\Omega(\beta^\prime
\sin\phi-\gamma^\prime\sin\beta\cos\phi)\right]+\mu_0r^2\Omega
\gamma^\prime\cos\beta\sin\phi(\beta^\prime\sin\phi-\gamma^\prime
\sin\beta\cos\phi)\nonumber\\
&&-\mu_0r\cos\phi(\beta^\prime\sin\phi-\gamma^\prime\sin\beta\cos\phi)
\left[(r\Omega)^\prime+(\beta^\prime\cos\phi+\gamma^\prime\sin\beta
\sin\phi)\partial_\zeta(v_{\phi1}+rv_{r1}\gamma^\prime\cos\beta)\right]
\big\rangle\,r\,{\rm
d}\zeta.
\end{eqnarray}
Again, it will be useful to consider the complex combination
\begin{eqnarray}
\lefteqn{I_2+{\rm i}I_3=\int{{1}\over{r^2}}(\partial_r+{\rm
i}\gamma^\prime\cos\beta)\big\langle{\rm e}^{{\rm
i}\phi}\left\{\rho_0r^4\Omega\zeta v_{r1}-{\rm
i}\rho_0r^3v_{r1}\left[v_{\theta1}+rv_{r1}(\beta^\prime\cos\phi+
\gamma^\prime\sin\beta\sin\phi)\right]\right.}&\nonumber\\
&&\left.\qquad+{\rm
i}\mu_0r^3(\beta^\prime\cos\phi+\gamma^\prime\sin\beta\sin\phi)
\partial_\zeta\left[v_{\theta1}+rv_{r1}(\beta^\prime\cos\phi+
\gamma^\prime\sin\beta\sin\phi)\right]-{\rm
i}\mu_0r^2\partial_\zeta v_{r1}\right.\nonumber\\
&&\left.\qquad-{\rm
i}\mu_0r^4\Omega(\beta^\prime\sin\phi-\gamma^\prime\sin\beta\cos\phi)
\right\}\big\rangle+\big\langle{\rm
e}^{{\rm i}\phi}\left\{-\rho_0(r^2\Omega)^\prime\zeta
v_{r1}\right.\nonumber\\
&&\left.\qquad-\rho_0(v_{\phi1}+rv_{r1}\gamma^\prime\cos\beta)
\left[v_{\theta1}+rv_{r1}(\beta^\prime\cos\phi+\gamma^\prime
\sin\beta\sin\phi)\right]+{\rm
i}\rho_0r\Omega\zeta(v_{\phi1}+rv_{r1}\gamma^\prime\sin\beta)
\right.\nonumber\\
&&\left.\qquad+{\rm
i}\rho_0rv_{r1}(v_{\phi1}+rv_{r1}\gamma^\prime\cos\beta)
(\beta^\prime\sin\phi-\gamma^\prime\sin\beta\cos\phi)-
{{\mu_0}\over{r}}\partial_\zeta(v_{\phi1}+rv_{r1}\gamma^\prime
\cos\beta)\right.\nonumber\\
&&\left.\qquad-{\rm
i}\mu_0r(\beta^\prime\sin\phi-\gamma^\prime\sin\beta\cos\phi)
\left[r\Omega^\prime+(\beta^\prime\cos\phi+\gamma^\prime\sin\beta
\sin\phi)\partial_\zeta(v_{\phi1}+rv_{r1}\gamma^\prime\cos\beta)
\right]\right\}\big\rangle\,r\,{\rm
d}\zeta.
\end{eqnarray}

\section{Separation of variables}

In order to proceed, it is helpful to make two simplifying, but
physically reasonable, assumptions.  First, the viscosity coefficients
are assumed to be locally proportional to the pressure, so that
\begin{eqnarray}
\mu&=&\alpha p/\Omega,\\
\mu_{\rm b}&=&\alpha_{\rm b}p/\Omega,
\end{eqnarray}
where the dimensionless coefficients $\alpha(r)$ and $\alpha_{\rm
b}(r)$ are prescribed functions of radius.  Secondly, the fluid is
assumed to be locally polytropic, with $\Gamma(r)>1$ being a
prescribed function of radius.  These assumptions allow the equations
of Set~A to be solved by separation of variables.  (An isothermal
assumption, $\Gamma=1$, would allow a similar simplification.)

Given that the pressure and density locally satisfy a polytropic
relation, $p=K\rho^\Gamma$, introduce the enthalpy
\begin{equation}
h=\int{{{\rm
d}p}\over{\rho}}=\left({{\Gamma}\over{\Gamma-1}}\right)K\rho^{\Gamma-1}=
\left({{\Gamma}\over{\Gamma-1}}\right){{p}\over{\rho}}.
\end{equation}
Then equations (\ref{a1}) and (\ref{a2}) may be replaced by
\begin{equation}
\left(\Omega\partial_\phi-{{v_{\theta1}}\over{r}}\partial_\zeta\right)h_0=
{{(\Gamma-1)h_0}\over{r}}\partial_\zeta v_{\theta1},
\end{equation}
and Set~A is reduced to a problem in four dependent variables
$\{h_0,v_{r1},v_{\theta1},(v_{\phi1}+rv_{r1}\gamma^\prime\cos\beta)\}$.
Note that $\psi=|\psi|\,{\rm e}^{{\rm i}\chi}$ occurs only in the
combinations
\begin{eqnarray}
r(\beta^\prime\cos\phi+\gamma^\prime\sin\beta\sin\phi)&=&|\psi|
\cos(\phi-\chi),\\
r(\beta^\prime\sin\phi-\gamma^\prime\sin\beta\cos\phi)&=&|\psi|
\sin(\phi-\chi).
\end{eqnarray}
The solution of Set~A is then of the form
\begin{eqnarray}
h_0&=&r^2\Omega^2\left[f_1(\phi-\chi)-{\textstyle{{1}\over{2}}}
f_2(\phi-\chi)\zeta^2\right],\\
v_{r1}&=&r\Omega f_3(\phi-\chi)\zeta,\\
v_{\theta1}&=&r\Omega f_4(\phi-\chi)\zeta,\\
v_{\phi1}+rv_{r1}\gamma^\prime\cos\beta&=&r\Omega f_5(\phi-\chi)\zeta.
\end{eqnarray}
where the dimensionless functions $f_1,\dots,f_5$ (whose parametric
dependence on $r$ and $T$ has been suppressed) satisfy the non-linear
ODEs
\begin{eqnarray}
f_1^\prime(\phi)&=&(\Gamma-1)f_4(\phi)f_1(\phi),\label{f1}\\
f_2^\prime(\phi)&=&(\Gamma+1)f_4(\phi)f_2(\phi),\label{f2}\\
f_3^\prime(\phi)&=&f_4(\phi)f_3(\phi)+2f_5(\phi)+\left[1+(\alpha_{\rm
b}+{\textstyle{{1}\over{3}}}\alpha)f_4(\phi)\right]f_2(\phi)|\psi|
\cos\phi-\alpha
f_2(\phi)f_3(\phi)(1+|\psi|^2\cos^2\phi)\nonumber\\ &&\qquad-\alpha
f_2(\phi)|\psi|\sin\phi,\label{f3}\\
f_4^\prime(\phi)&=&-f_3^\prime(\phi)|\psi|\cos\phi+2f_3(\phi)|\psi|
\sin\phi+f_4(\phi)\left[f_4(\phi)+f_3(\phi)|\psi|\cos\phi\right]+
1-\left[1+(\alpha_{\rm
b}+{\textstyle{{1}\over{3}}}\alpha)f_4(\phi)\right]f_2(\phi)\nonumber\\
&&\qquad-\alpha
f_2(\phi)\left[f_4(\phi)+f_3(\phi)|\psi|\cos\phi\right]
(1+|\psi|^2\cos^2\phi)+\alpha
f_2(\phi)|\psi|^2\cos\phi\sin\phi,\label{f4}\\
f_5^\prime(\phi)&=&f_4(\phi)f_5(\phi)-{\textstyle{{1}\over{2}}}
\tilde\kappa^2f_3(\phi)-\alpha
f_2(\phi)f_5(\phi)(1+|\psi|^2\cos^2\phi)+{\textstyle{{1}\over{2}}}
(4-\tilde\kappa^2)\alpha
f_2(\phi)|\psi|\cos\phi,\label{f5}
\end{eqnarray}
subject to periodic boundary conditions $f_n(2\pi)=f_n(0)$.  There are
five dimensionless parameters
$\{|\psi|,\tilde\kappa^2,\Gamma,\alpha,\alpha_{\rm b}\}$.  Note that
equation (\ref{f1}) is decoupled, and that $f_1$ admits an arbitrary
multiplicative constant which allows the surface density to be fitted.
The (upper) surface of the disc is given at leading order by
\begin{equation}
\zeta=\zeta_{\rm
s}(\phi)=\left[{{2f_1(\phi-\chi)}\over{f_2(\phi-\chi)}}\right]^{1/2}.
\end{equation}

The expressions $I_n$ of Section~4.5 may now be simplified.  The
vertical integrals introduce the quantities $\tilde{\sI}$ and
$\tilde{\sV}$ which are related by
\begin{equation}
\tilde{\sV}=\alpha f_2(\phi-\chi)\Omega\tilde{\sI}.
\end{equation}
This follows from the definition of $\tilde{\sV}$ after an integration
by parts.  Then equations (\ref{I1}) and (\ref{new1}) can be
identified provided that
\begin{eqnarray}
Q_1&=&\big\langle
f_6\left[-{\textstyle{{1}\over{2}}}(4-\tilde\kappa^2)\alpha
f_2-f_3f_5+\alpha f_2f_5|\psi|\cos\phi\right]\big\rangle,\label{q1}\\
Q_2&=&{{1}\over{|\psi|^2}}\big\langle
f_6\left[(f_4+f_3|\psi|\cos\phi)(1+f_3|\psi|\sin\phi)+\alpha
f_2f_3|\psi|\sin\phi-\alpha
f_2(f_4+f_3|\psi|\cos\phi)|\psi|^2\cos\phi\sin\phi\right.\nonumber\\
&&\left.\qquad+\alpha
f_2|\psi|^2\sin^2\phi\right]\big\rangle,\label{q2}
\end{eqnarray}
in which $f_n$ stands for $f_n(\phi)$, and
\begin{equation}
f_6(\phi-\chi)=\tilde{\sI}/\sI
\end{equation}
contains the azimuthal dependence of $\tilde{\sI}$.  To evaluate this,
note that
\begin{equation}
\int(f_1-{\textstyle{{1}\over{2}}}f_2\zeta^2)^{1/(\Gamma-1)}\zeta^2\,{\rm
d}\zeta\propto f_1^{1/(\Gamma-1)}\left({{f_1}\over{f_2}}\right)^{3/2},
\end{equation}
and so
\begin{equation}
f_6(\phi)=[f_1(\phi)]^{1/(\Gamma-1)}\left[{{f_1(\phi)}
\over{f_2(\phi)}}\right]^{3/2}\bigg/\bigg\langle
f_1^{1/(\Gamma-1)}\left({{f_1}\over{f_2}}\right)^{3/2}\bigg\rangle.
\end{equation}
Then it follows from equations (\ref{f1}) and (\ref{f2}) that
\begin{equation}
f_6^\prime(\phi)=-2f_4(\phi)f_6(\phi),\label{f6}
\end{equation}
while the definition of $f_6$ requires $\langle f_6\rangle=1$.  Note
that, for any function $F$,
\begin{equation}
\langle{\rm e}^{{\rm i}\phi}F(\phi-\chi)\rangle=\langle{\rm e}^{{\rm
i}(\phi+\chi)}F(\phi)\rangle={{\psi}\over{|\psi|}}\langle{\rm e}^{{\rm
i}\phi}F(\phi)\rangle.
\end{equation}
Thus $I_2+{\rm i}I_3$ can be identified with $X+{\rm i}Y$, provided
that
\begin{eqnarray}
Q_1&=&{1\over{|\psi|}}\big\langle{\rm e}^{{\rm
i}\phi}f_6\left\{-{\textstyle{{1}\over{2}}}\tilde\kappa^2
f_3-f_5(f_4+f_3|\psi|\cos\phi)+{\rm i}f_5+{\rm
i}f_3f_5|\psi|\sin\phi-\alpha f_2f_5\right.\nonumber\label{q1again}\\
&&\left.\qquad-{\rm i}\alpha
f_2[-{\textstyle{{1}\over{2}}}(4-\tilde\kappa^2)+
f_5|\psi|\cos\phi]|\psi|\sin\phi\right\}\big\rangle\\
Q_4&=&{1\over{|\psi|}}\big\langle{\rm e}^{{\rm
i}\phi}f_6\left[f_3-{\rm i}f_3(f_4+f_3|\psi|\cos\phi)+{\rm i}\alpha
f_2(f_4+f_3|\psi|\cos\phi)|\psi|\cos\phi-{\rm i}\alpha f_2f_3-{\rm
i}\alpha f_2|\psi|\sin\phi\right]\big\rangle,\label{q4}
\end{eqnarray}
These correspondences are equivalent: equations (\ref{q1}) and
(\ref{q1again}), and equations (\ref{q2}) and (\ref{q4}), can be shown
to agree by using equations (\ref{f5}) and (\ref{f6}).  Therefore
equations (\ref{q1}) and (\ref{q4}) for $Q_1$ and $Q_4$ are sufficient
and will be taken as definitive.  This completes the demonstration
that the fully non-linear problem can be reduced to one-dimensional
conservation equations for mass and angular momentum as anticipated in
Section~3.2.

In the inviscid case $\alpha=\alpha_{\rm b}=0$, the functions $f_1$,
$f_2$, $f_5$ and $f_6$ are even, while $f_3$ and $f_4$ are odd.  Then
$Q_1=Q_2=0$, i.e. $Q_4$ is purely imaginary.

\section{Evaluation of the coefficients}

In principle, equations (\ref{f1})--(\ref{f5}) and (\ref{f6}) can be
expanded in powers of $|\psi|$ to determine the dynamics to any
desired order.  In the Appendix, truncated Taylor series for the
coefficients $Q_1$, $Q_2$ and $Q_3$ are developed in this way.  This
generalizes the linear theory of Papaloizou \& Pringle (1983) to allow
for an arbitrary rotation law, and extends it into the weakly
non-linear domain.  These series can be used to estimate at which
amplitude, and in what way, the linear theory breaks down.  This
behaviour is interpreted further in Section~7.  The expansion fails
only when both $\tilde\kappa^2\to1$ and $\alpha\to0$; this is the
previously identified resonant case, which cannot be described using
this method.

It is also straightforward to integrate equations
(\ref{f2})--(\ref{f5}) and (\ref{f6}) numerically, imposing periodic
boundary conditions and the normalization condition $\langle
f_6\rangle=1$.  The coefficients $Q_1$, $Q_2$ and $Q_3$ can then be
determined directly for any amplitude of warp, resulting in a fully
non-linear theory.  Two important cases have been investigated in this
way.

\subsection{Inviscid, non-Keplerian disc}

The first case is that of an inviscid disc with $\Gamma=5/3$ and
$\alpha=\alpha_{\rm b}=0$.  The remaining parameters are $|\psi|$ and
$\tilde\kappa^2$.  A contour plot of the only non-vanishing
coefficient, $Q_3$, is shown in Fig.~2.  For reasonably small values
of $|\psi|$ there is good agreement with the truncated Taylor series.
Where $\tilde\kappa^2>1$, $Q_3$ is negative and the solution can be
continued to large values of $|\psi|$, although the disc becomes
highly non-axisymmetric in this limit.  Where $\tilde\kappa^2<1$,
$Q_3$ is positive and the solution exists only for sufficiently small
$|\psi|$.  The solution terminates when $f_2$ becomes zero at some
point, with the result that the disc can no longer contain itself
vertically and ruptures.

\begin{figure}
\centerline{\epsfbox{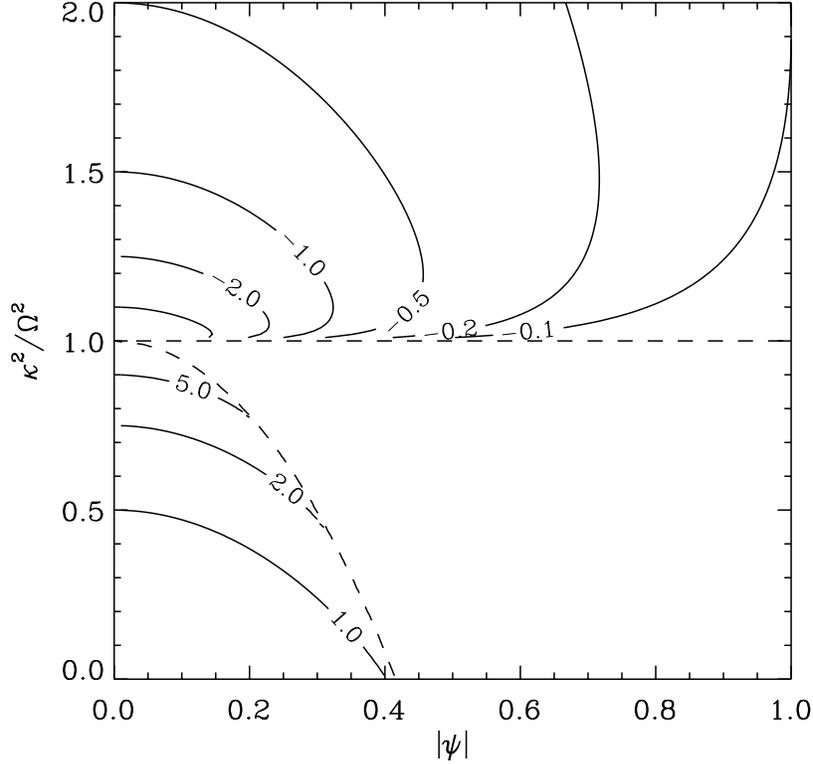}}
\caption{Contour plot of the coefficient $Q_3$ for an inviscid disc
with $\Gamma=5/3$.  The horizontal coordinate is the amplitude of the
warp.  The vertical coordinate is the square of the dimensionless
epicyclic frequency.  The dashed lines bound the region in which a
solution is found.}
\end{figure}

\subsection{Viscous, Keplerian disc}

The second case is that of a viscous, Keplerian disc with
$\tilde\kappa^2=1$, $\Gamma=5/3$ and $\alpha_{\rm b}=0$.  The
remaining parameters are $|\psi|$ and $\alpha$.  Contour plots of the
three coefficients are shown in Figs~3, 4 and 5.  Again, for
reasonably small values of $|\psi|$ there is good agreement with the
truncated Taylor series, except when $\alpha$ is small.  The solution
can be continued to large values of $|\psi|$ for any value of
$\alpha$, and again the disc becomes highly non-axisymmetric in this
limit.  There are several features to note.  First, as predicted by
the Taylor series, $Q_1$ increases with increasing $|\psi|$ and even
becomes positive for sufficiently small $\alpha$.  Secondly, $Q_2$ has
a strong peak at $|\psi|=\alpha=0$ where the resonance is located, but
the resonance does not extend to large values of $|\psi|$.  Thirdly,
$Q_3$ is typically much smaller in magnitude than $Q_2$.

\begin{figure}
\centerline{\epsfbox{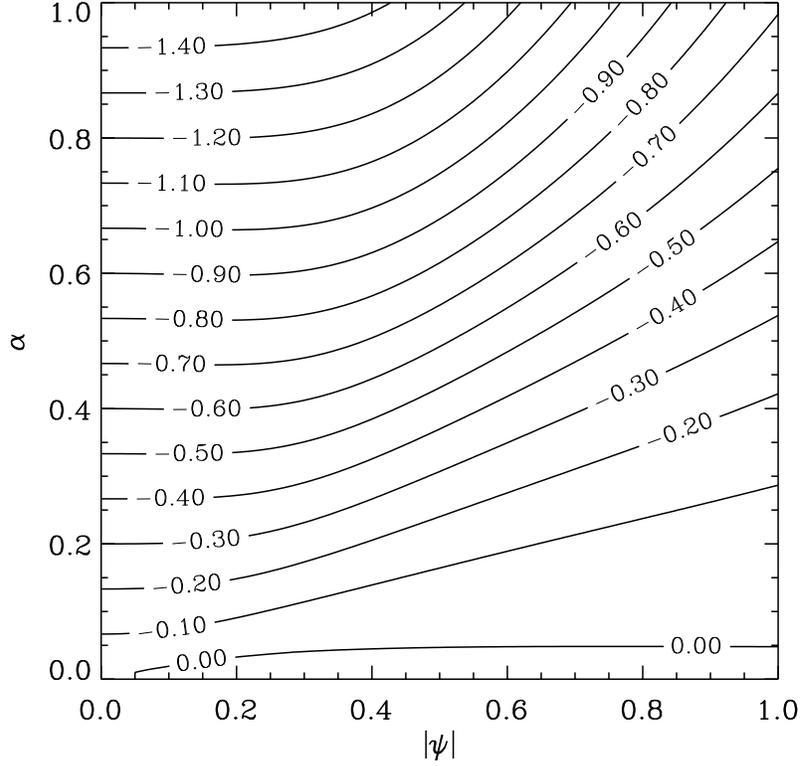}}
\caption{Contour plot of the coefficient $Q_1$ for a viscous,
Keplerian disc with $\Gamma=5/3$ and $\alpha_{\rm b}=0$.  The
horizontal coordinate is the amplitude of the warp.  The vertical
coordinate is the dimensionless viscosity parameter.}
\end{figure}

\begin{figure}
\centerline{\epsfbox{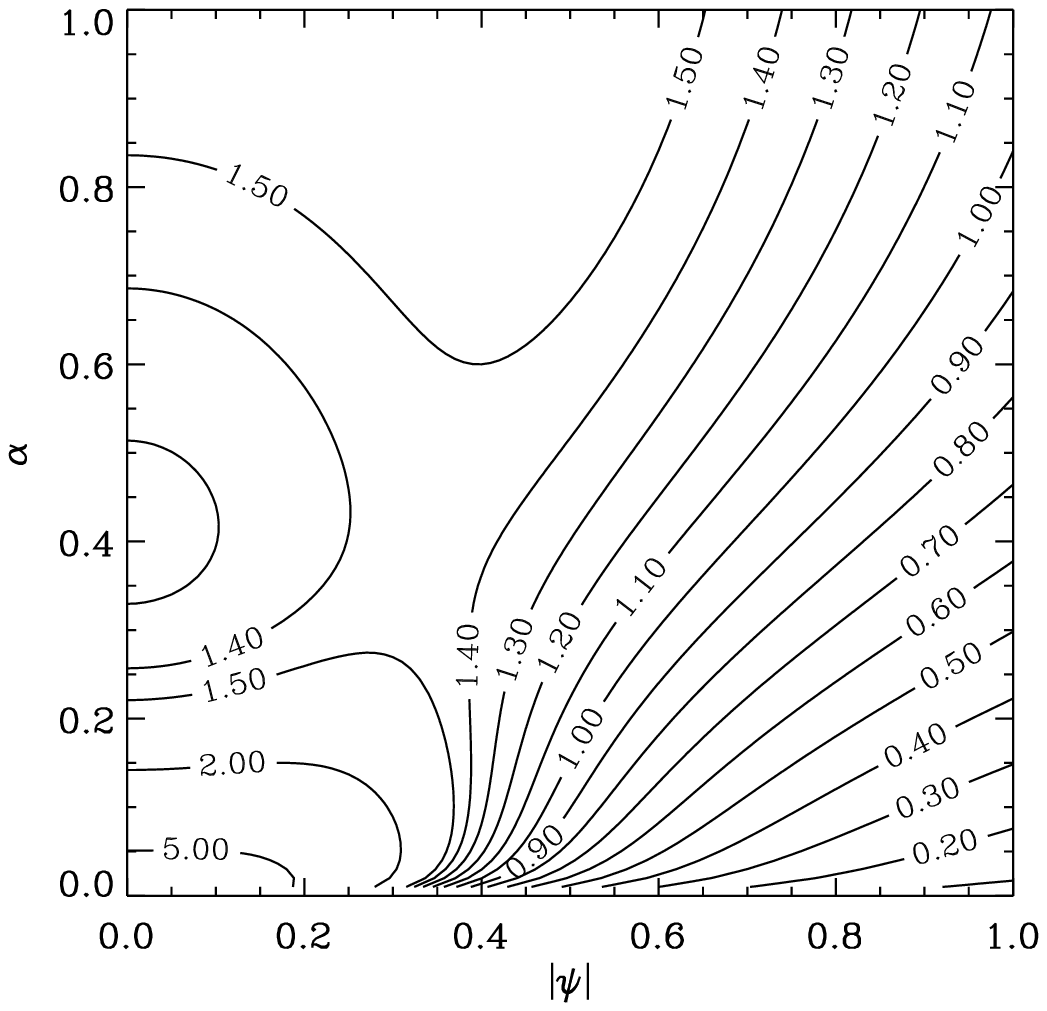}}
\caption{Contour plot of the coefficient $Q_2$ for a viscous,
Keplerian disc with $\Gamma=5/3$ and $\alpha_{\rm b}=0$.}
\end{figure}

\begin{figure}
\centerline{\epsfbox{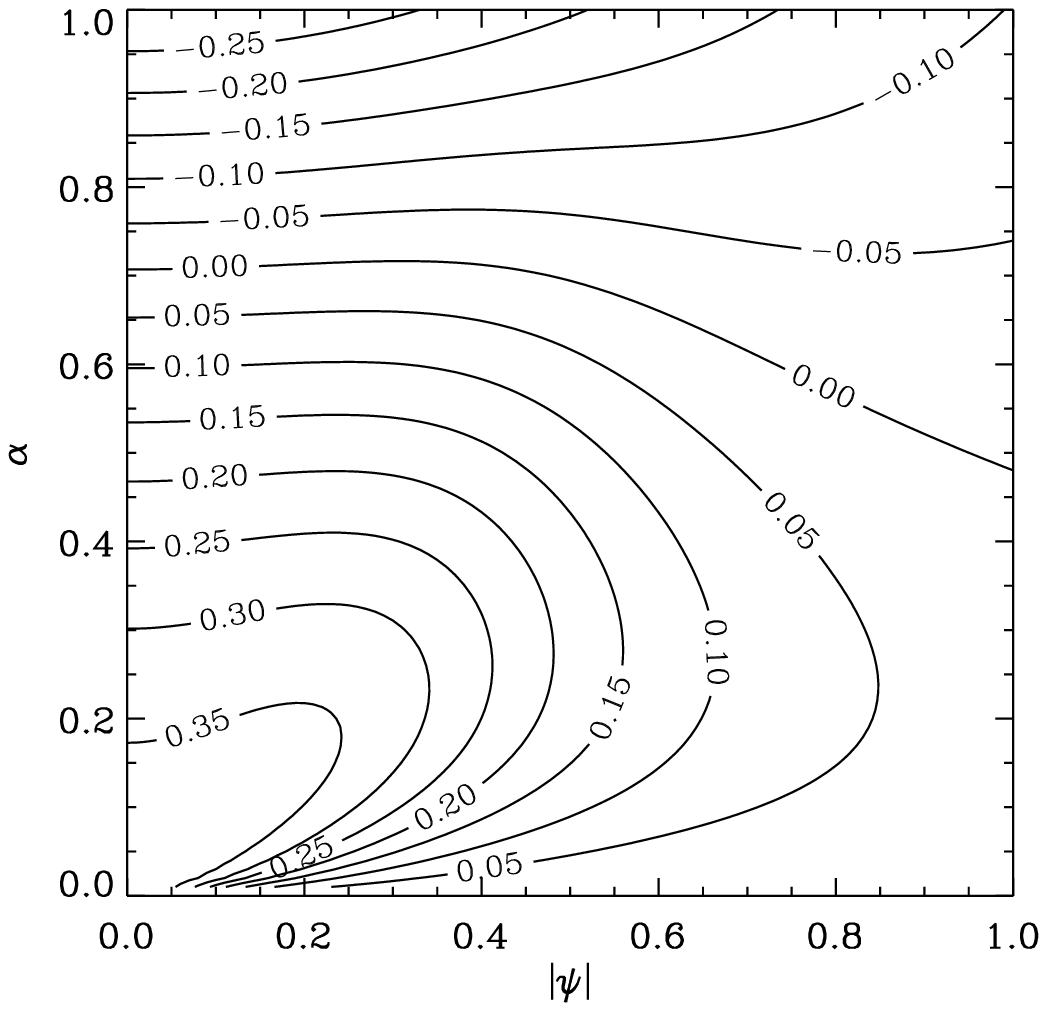}}
\caption{Contour plot of the coefficient $Q_3$ for a viscous,
Keplerian disc with $\Gamma=5/3$ and $\alpha_{\rm b}=0$.}
\end{figure}

\section{Summary and interpretation}

\subsection{General remarks}

In this paper, the non-linear fluid dynamics of a warped accretion
disc has been investigated by developing a theory of fully non-linear
bending waves for a thin, viscous disc in a spherically symmetric
external potential.  It has been found that the dynamics is described
by an equation for the surface density,
\begin{equation}
{{\partial\Sigma}\over{\partial t}}+{{1}\over{r}}{{\partial}\over
{\partial r}}(r\Sigma\bar v_r)=0,
\end{equation}
and an equation for the angular momentum,
\begin{equation}
{{\partial}\over{\partial t}}(\Sigma
r^2\Omega{\bmath\ell})+{{1}\over{r}}{{\partial}\over{\partial
r}}(\Sigma\bar
v_rr^3\Omega{\bmath\ell})={{1}\over{r}}{{\partial}\over{\partial
r}}\left(Q_1\sI
r^2\Omega^2{\bmath\ell}\right)+{{1}\over{r}}{{\partial}\over{\partial
r}}\left(Q_2\sI r^3\Omega^2{{\partial{\bmath\ell}}\over{\partial
r}}\right)+{{1}\over{r}}{{\partial}\over{\partial r}}\left(Q_3\sI
r^3\Omega^2{\bmath\ell}\times{{\partial{\bmath\ell}}\over{\partial
r}}\right)+{\bmath T}.
\end{equation}
Here $\Sigma(r,t)$ is the surface density, $\bar v_r(r,t)$ is the mean
radial velocity, $\Omega(r)$ is the orbital angular velocity,
${\bmath\ell}(r,t)$ is the tilt vector, $\sI(r,t)$ is the azimuthally
averaged second vertical moment of the density [cf. equation
(\ref{i})], and a new term ${\bmath T}$ represents any additional
torque due to self-gravitation, radiation forces, tidal forcing, etc.
The dimensionless coefficients $Q_1$, $Q_2$ and $Q_3$ depend on the
rotation law and the shear viscosity, and also, in the non-linear
theory, on the adiabatic exponent, the bulk viscosity and the
amplitude of the warp.  They have been determined both analytically,
as truncated Taylor series in the amplitude of the warp, and
numerically, by solving a set of ordinary differential equations.  The
angular momentum equation can be decomposed into an equation for the
component parallel to ${\bmath\ell}$,
\begin{equation}
\Sigma\bar v_r{{{\rm d}(r^2\Omega)}\over{{\rm
d}r}}={{1}\over{r}}{{\partial}\over{\partial r}}\left(Q_1\sI
r^2\Omega^2\right)-Q_2\sI
r^2\Omega^2\left|{{\partial{\bmath\ell}}\over{\partial
r}}\right|^2+{\bmath\ell}\cdot{\bmath T}
\end{equation}
and an equation for the tilt vector,
\begin{equation}
\Sigma r^2\Omega\left({{\partial{\bmath\ell}}\over{\partial t}}+\bar
v_r{{\partial{\bmath\ell}}\over{\partial r}}\right)=Q_1\sI
r\Omega^2{{\partial{\bmath\ell}}\over{\partial
r}}+{{1}\over{r}}{{\partial}\over{\partial r}}\left(Q_2\sI
r^3\Omega^2{{\partial{\bmath\ell}}\over{\partial r}}\right)+Q_2\sI
r^2\Omega^2\left|{{\partial{\bmath\ell}}\over{\partial
r}}\right|^2{\bmath\ell}+{{1}\over{r}}{{\partial}\over{\partial
r}}\left(Q_3\sI
r^3\Omega^2{\bmath\ell}\times{{\partial{\bmath\ell}}\over{\partial
r}}\right)-{\bmath\ell}\times({\bmath\ell}\times{\bmath T}).
\end{equation}
This scheme is a generalization of the form proposed by Pringle
(1992), and is equally suitable for numerical implementation.  The
internal torques between one ring in the disc and its neighbours are
of three kinds.  Coefficient $Q_1$ represents a torque tending to spin
up (or down) the ring.  This would be the usual viscous torque
proportional to ${\rm d}\Omega/{\rm d}r$ in a flat disc, but in a
warped disc there is an additional contribution, not proportional to
${\rm d}\Omega/{\rm d}r$, due to a correlation between the radial and
azimuthal velocities induced by the warp; this vanishes in an inviscid
disc because the radial and azimuthal velocities are perfectly out of
phase.  Coefficient $Q_2$ represents a torque tending to align the
ring with its neighbours, which acts to flatten the disc; this also
vanishes in an inviscid disc.\footnote{It is to be expected that $Q_2$
will normally be positive, since it represents a diffusion
coefficient.  The linear theory (see the Appendix) suggests that this
is true except possibly for unphysical rotation laws
($\tilde\kappa^2>8$).}  Coefficient $Q_3$ represents a torque tending
to make the ring precess if it is misaligned with its neighbours; this
leads to the dispersive wave-like propagation of the warp.  It is
likely that this form of angular momentum equation applies in much
more general circumstances than those considered in this paper.
However, the present analysis provides a consistent evaluation of the
three coefficients under a number of simplifying assumptions.

There are essentially four different assumptions or approximations
involved in this analysis.  The first assumption is that the disc is
thin ($H/r\ll1$), which is essential for the asymptotic expansions in
Section~4.  The second assumption is that the fluid obeys the
compressible Navier-Stokes equation even though, in reality, the
turbulent stresses in an accretion disc are likely to be anisotropic
and may not be purely viscous.  The third assumption is that the fluid
is locally polytropic and has viscosity coefficients locally
proportional to the pressure.  These are the simplest possible closure
relations in the absence of a full treatment of thermal and radiative
physics, and greatly simplify the problem by allowing separation of
variables (Section~5).  In a fully self-consistent treatment, the
thickness of the disc, and therefore the relation between $\sI$ and
$\Sigma$, would be determined by the local viscous dissipation of
energy.  The final assumption is that the conditions for resonance,
\begin{equation}
\left|{{\Omega^2-\kappa^2}\over{\Omega^2}}\right|\la
H/r\qquad\hbox{and}\qquad\alpha\la H/r
\end{equation}
are not simultaneously satisfied.  (It has been shown in Section~6.2
that a further condition for resonance is that the amplitude of the
warp be small.)  The asymptotic expansions formally break down as the
resonance is approached.  The nature of the approximation is as
follows.  In equations (\ref{a3}) and (\ref{a5}) governing the
horizontal velocities induced by the warp, the time-derivatives of the
velocities (in the inertial frame) do not appear because those terms
are of higher order in the assumed ordering scheme.  The horizontal
velocities are in a `geostrophic' balance in which the driving forces
are matched instantaneously by inertial and viscous forces, and for
this reason they do not constitute additional dynamical degrees of
freedom.  In the resonant case, this approximation breaks down and the
time-derivatives must be restored.  This results in a different
ordering scheme and makes the problem hyperbolic rather than parabolic
(Papaloizou \& Lin 1994).  The non-linear dynamics of the resonant
case is likely to be very different.

Two important cases merit further interpretation.

\subsection{Inviscid, non-Keplerian disc}

In the inviscid case, the radial velocity vanishes, the surface
density is independent of time, and the dynamics is described by an
equation for the tilt vector of the form
\begin{equation}
\Sigma r^2\Omega{{\partial{\bmath\ell}}\over{\partial
t}}={{1}\over{r}}{{\partial}\over{\partial r}}\left(Q_3\sI
r^3\Omega^2{\bmath\ell}\times{{\partial{\bmath\ell}}\over{\partial
r}}\right),
\end{equation}
where $Q_3$ is plotted in Fig.~2 for the case $\Gamma=5/3$.  It has
the Taylor series (see the Appendix)
\begin{equation}
Q_3={{1}\over{2(1-\tilde\kappa^2)}}+{{(6+\Gamma)}\over{4(3-\Gamma)
(1-\tilde\kappa^2)^2}}|\psi|^2+O(|\psi|^4),
\end{equation}
where
\begin{equation}
|\psi|=r\left|{{\partial{\bmath\ell}}\over{\partial r}}\right|
\end{equation} is the amplitude of the warp.  Equivalently,
\begin{equation}
\Sigma r^2\Omega{{\partial{\bmath\ell}}\over{\partial
t}}={{1}\over{r}}{{\partial}\over{\partial r}}\left\{\left[{{\sI
r^3\Omega^4}\over{2(\Omega^2-\kappa^2)}}+{{(6+\Gamma)\sI
r^5\Omega^6}\over{4(3-\Gamma)(\Omega^2-\kappa^2)^2}}\left|
{{\partial{\bmath\ell}}\over{\partial
r}}\right|^2+O\left(\left|{{\partial{\bmath\ell}}\over{\partial
r}}\right|^4\right)\right]{\bmath\ell}\times{{\partial{\bmath\ell}}
\over{\partial
r}}\right\}.
\end{equation}
To make the connection with the linear theory of bending waves,
suppose that the disc is initially flat, with ${\bmath\ell}={\bmath
e}_z$, and is then perturbed weakly, with $|\ell_x|,|\ell_y|\ll1$.
Then
\begin{equation}
\Sigma r^2\Omega{{\partial W}\over{\partial
t}}={{1}\over{r}}{{\partial}\over{\partial r}}\left[{{\sI
r^3\Omega^4}\over{2(\Omega^2-\kappa^2)}}\left({\rm i}{{\partial
W}\over{\partial r}}+{\textstyle{{1}\over{2}}}{\rm
i}{{\partial|W|^2}\over{\partial r}}W-{\textstyle{{1}\over{2}}}{\rm
i}|W|^2{{\partial W}\over{\partial r}}\right)+{{(6+\Gamma)\sI
r^5\Omega^6}\over{4(3-\Gamma)(\Omega^2-\kappa^2)^2}}{\rm
i}|W^\prime|^2{{\partial W}\over{\partial r}}+O(W^5)\right],\label{w}
\end{equation}
where $W=\ell_x+{\rm i}\ell_y$.  This is a Schr\"odinger-type wave
equation with non-linear dispersion.  The linearized form of this
equation,
\begin{equation}
\Sigma r^2\Omega{{\partial W}\over{\partial
t}}={{1}\over{r}}{{\partial}\over{\partial r}}\left[{{\sI
r^3\Omega^4}\over{2(\Omega^2-\kappa^2)}}\left({\rm i}{{\partial
W}\over{\partial r}}\right)\right],\label{wlinear}
\end{equation}
may be compared with equation (12) of Papaloizou \& Lin (1994).
Taking the low-frequency, non-resonant limit of their equation,
setting the precession frequency to zero for a spherical potential,
restoring the time-derivative and identifying $W\leftrightarrow g^*$,
one obtains
\begin{equation}
\Sigma r^2\Omega{{\partial W}\over{\partial
t}}=r^2\Omega^2{{\partial}\over{\partial
r}}\left[{{\sI\Omega^2}\over{2(\Omega^2-\kappa^2)}}\left({\rm
i}{{\partial W}\over{\partial r}}\right)\right].
\end{equation}
The only discrepancy concerns whether a factor $r^3\Omega^2$ should
appear inside or outside the radial derivative.  This is of little
significance for nearly Keplerian discs in which $r^3\Omega^2$ is
approximately constant, and may be attributable to a failure of the
approximations adopted in the passage from equation (11) to equation
(12) of that paper.  It is evident that equation (\ref{wlinear}) is of
the correct form to ensure the conservation of angular momentum.

In a short-wavelength (WKB) limit, equation (\ref{w}) has wave-like
solutions of the form
\begin{equation}
W(r,t)=\tilde W(r)\exp\left[-{\rm i}\omega t+{\rm i}\int k(r)\,{\rm
d}r\right],
\end{equation}
where $\omega$ is the frequency, $k(r)$ is the radial wavenumber
(satisfying $|kr|\gg1$) and $\tilde W(r)$ is a slowly varying
function.  The non-linear WKB dispersion relation is then
\begin{equation}
{{\omega}\over{\Omega}}=\pm{{1}\over{2}}\left({{\Omega^2}\over
{\Omega^2-\kappa^2}}\right){{\sI
k^2}\over{\Sigma}}\left[1+{{(6+\Gamma)}\over{2(3-\Gamma)}}
\left({{\Omega^2}\over{\Omega^2-\kappa^2}}\right)k^2r^2|\tilde
W|^2+O(|\tilde W|^4)\right],
\end{equation}
the $\pm$ arising since the complex-conjugate solution is equally
valid but is physically distinct.  In the linear approximation, this
can be shown to agree with the analytic WKB dispersion relation for
isothermal discs found by Lubow \& Pringle (1993).  To make the
connection, note that their $K_x$ is our $kH$, where $H$ is the
isothermal scale height of the disc, while their $F$ is our
$(\omega\pm\Omega)/\Omega$ for an $m=\mp1$ mode.  Expanding their
equation (54) about the point $K_x=0$, $F=\pm1$, and assuming a
non-Keplerian disc, one obtains for the `$n=0$' mode
\begin{equation}
{{\omega}\over{\Omega}}=\pm{{1}\over{2}}\left({{\Omega^2}\over
{\Omega^2-\kappa^2}}\right)(kH)^2+O\left((kH)^4\right).
\end{equation}
For an isothermal disc $\sI=\Sigma H^2$ and the two agree.

The interpretation of the non-linear term in the dispersion relation
is that the dispersion of waves either increases or decreases as the
amplitude increases, depending on the signs of $(\Omega^2-\kappa^2)$
and of $(3-\Gamma)$.  The cubic non-linearity arises through a
three-mode coupling involving (i) the `tilt' mode or f mode ($m=1$, of
odd symmetry), consisting locally of a uniform vertical translation of
the disc, (ii) an inertial or r mode ($m=1$, of odd symmetry),
consisting of a horizontal epicyclic motion proportional to $\zeta$,
and (iii) an acoustic or p mode ($m=2$, of even symmetry), consisting
of a vertical motion proportional to $\zeta$.  The magnitude of the
coupling depends on the adiabatic exponent because of the compressive
nature of the p mode, which has a natural frequency of
$(1+\Gamma)^{1/2}\Omega$ in the comoving frame.  Indeed, in the
unlikely case $\Gamma=3$, the p mode would be driven at resonance and
a large response would result.

Fig.~2 shows that, when $\Gamma=5/3$, the non-linear behaviour is very
different depending on the sign of $(\Omega^2-\kappa^2)$.  If
$\kappa^2>\Omega^2$ the disc can support warps of large amplitude and
the dispersion coefficient becomes smaller in magnitude as the
amplitude increases.  However, if $\kappa^2<\Omega^2$, the dispersion
coefficient increases with increasing amplitude and the disc
eventually ruptures.

\subsection{Viscous, Keplerian disc}

In the viscous, Keplerian case, the three coefficients are plotted in
Figs~3, 4 and 5 (with $\Gamma=5/3$ and $\alpha_{\rm b}=0$).  They have
Taylor series (see the Appendix)
\begin{eqnarray}
Q_1&=&-{{3\alpha}\over{2}}+\left[{{1-17\alpha^2+21\alpha^4}\over
{4\alpha(4+\alpha^2)}}\right]|\psi|^2+O(|\psi|^4),\\ Q_2+{\rm
i}Q_3&=&{{1+2{\rm i}\alpha+6\alpha^2}\over{2\alpha(2+{\rm
i}\alpha)}}+\left\{{{\tilde a+\tilde b\Gamma+\tilde
c\left[\Gamma-2{\rm i}(\alpha_{\rm
b}+{\textstyle{{4}\over{3}}}\alpha)\right]}\over{4\alpha\left[
(3-\Gamma)+2{\rm
i}(\alpha_{\rm b}+{\textstyle{{4}\over{3}}}\alpha)\right](2{\rm
i}-\alpha)^3(2{\rm i}+\alpha)}}\right\}|\psi|^2+O(|\psi|^4),
\end{eqnarray}
where
\begin{eqnarray}
\tilde a&=&12+115{\rm i}\alpha+49\alpha^2+109{\rm
i}\alpha^3-408\alpha^4-41{\rm i}\alpha^5+2\alpha^6+4{\rm i}\alpha^7,\\
\tilde b&=&18+87{\rm i}\alpha-87\alpha^2-24\alpha^4-6{\rm
i}\alpha^5,\\ \tilde c&=&12+25{\rm i}\alpha-36\alpha^2-11{\rm
i}\alpha^3+140\alpha^4+21{\rm i}\alpha^5+2\alpha^6.
\end{eqnarray}
In the important limit $\alpha\ll1$, $\alpha_{\rm b}\ll1$ (but
requiring $\alpha\ga H/r$ to avoid the resonant case),
\begin{eqnarray}
Q_1&\approx&-{{3\alpha}\over{2}}+{{1}\over{16\alpha}}|\psi|^2+O(|\psi|^4),\\
Q_2&\approx&{{1}\over{4\alpha}}+O(|\psi|^2),\\
Q_3&\approx&{{3}\over{8}}+O(|\psi|^2).
\end{eqnarray}
This correctly predicts the behaviour seen in Fig.~3, in which the
usual viscous torque parallel to ${\bmath\ell}$ can be reversed by a
warp of amplitude $|\psi|\ga\sqrt{24}\alpha$.  This occurs because the
usual viscous torque is small (proportional to $\alpha$) and is easily
overwhelmed by the torque resulting from the correlation of the radial
and azimuthal velocities, since these are almost resonantly driven.
It also shows that the diffusion of the warp is more important than
any dispersive wave propagation in this limit.  Unfortunately, the
truncated Taylor series for $Q_2$ is inaccurate even for reasonably
small values of $|\psi|$ in this limit, and does not predict that it
ultimately decreases with increasing $|\psi|$, as seen in Fig.~4.
Therefore a numerical solution is required in this case.

To the extent that the dispersion coefficient $Q_3$ can be neglected,
the original equations of Pringle (1992) are formally valid, but with
the following caveats concerning the viscosities $\nu_1$ and $\nu_2$.
For small-amplitude warps, the approximations
\begin{equation}
\nu_1\approx\bar\nu
\end{equation}
and
\begin{equation}
\nu_2\approx\left[{{2(1+7\alpha^2)}\over{\alpha^2(4+\alpha^2)}}\right]
\bar\nu\approx{{\bar\nu}\over{2\alpha^2}}\qquad\hbox{for $\alpha\ll1$}
\end{equation}
hold (cf. Papaloizou \& Pringle 1983).  These are the approximations
relevant for deciding whether a flat disc is unstable to the
radiation-driven instability (Pringle 1996).  The fact that $\nu_2$ is
potentially much larger than $\nu_1$ means that the radius outside
which the instability sets in is likely to be much larger than would
be estimated on the assumption that $\nu_2=\nu_1$.  If the instability
does proceed, then $\nu_1$ and $\nu_2$ are subject to non-linear
corrections depending on the amplitude of the warp.  In the case
investigated in this paper, both $\nu_1$ and $\nu_2$ typically
decrease with increasing amplitude, with $\nu_1$ even becoming
negative if $\alpha$ is sufficiently small.  Although this means that
the usual accretion torque is reversed, the disc so strongly resists
being warped in this limit that this situation may not persist for
long unless the warp is strongly forced.

\subsection{Outlook}

This work has shown that, with the exception of discs that are both
accurately Keplerian and almost inviscid, the dynamics of a warped
accretion disc can be reduced to simple one-dimensional conservation
equations for mass and angular momentum even if the warp is
non-linear.  This generally supports the approach adopted by Pringle
(1992), although, for completeness, it requires an additional type of
internal torque to be included.  Based on the assumption of an
isotropic underlying viscous process, it also predicts the values of
the coefficients in the equations and their variation with local
parameters of the disc and with the amplitude of the warp.

Some features of the behaviour of the system can be estimated by
inspection of the equations and the variation of the coefficients.
Non-linear effects are strongest in the case of most interest, that of
a Keplerian disc with $H/r\la\alpha\ll1$.  However, the detailed
application of this work to astrophysical situations properly requires
a numerical solution of the equations.  As shown by Pringle (1992 et
seq.), the reduction of the problem to one-dimensional equations which
do not require the fast orbital time-scale to be followed offers very
significant practical advantages over three-dimensional numerical
simulations.

At the same time, there are two important ways in which this analysis
should be improved in future.  First, a better analytical modelling of
the turbulent stress tensor in accretion discs is desirable.  It
should be possible, using local numerical simulations, to measure the
response of magnetohydrodynamic turbulence to imposed motions such as
those induced by a warp (Torkelsson et al., in preparation) and to use
this to calibrate an analytical model.  Secondly, the effect of
increased dissipation caused by the warp on the thickness of the disc,
and therefore on the torques between neighbouring rings, ought to be
taken into account.  Unfortunately, to do this properly requires the
solution of an energy equation including radiative transport, which is
unlikely to be amenable to the technique of separation of variables
used in this paper.  However, rather than invalidating the form of the
equations derived here, these developments would be expected only to
provided improved values for the coefficients.

\section*{Acknowledgments}

I thank Jim Pringle for many helpful discussions.  I acknowledge the
hospitality of the Isaac Newton Institute during the programme
`Dynamics of Astrophysical Discs', where I benefited from discussions
with many participants, and where much of this work was completed.

\appendix

\section{Truncated non-linear equations}

The aim of this section is to derive truncated Taylor series for the
coefficients $Q_1$ and $Q_4=Q_2+{\rm i}Q_3$, which will provide an
equation for the tilt vector correct to cubic order.  To achieve this,
introduce the expansions
\begin{eqnarray}
f_1(\phi)&=&f_{10}+|\psi|^2f_{12}(\phi)+O(|\psi|^4),\\
f_2(\phi)&=&f_{20}+|\psi|^2f_{22}(\phi)+O(|\psi|^4),\\
f_3(\phi)&=&|\psi|f_{31}(\phi)+|\psi|^3f_{33}(\phi)+O(|\psi|^5),\\
f_4(\phi)&=&|\psi|^2f_{42}(\phi)+O(|\psi|^4),\\
f_5(\phi)&=&|\psi|f_{51}(\phi)+|\psi|^3f_{53}(\phi)+O(|\psi|^5),\\
f_6(\phi)&=&f_{60}+|\psi|^2f_{62}(\phi)+O(|\psi|^4).
\end{eqnarray}
Note that each function is either even or odd in $|\psi|$, and that
all terms with the scaling $|\psi|^0$ are axisymmetric, since they
represent an unwarped disc.

\subsection{Zeroth-order solution}

At zeroth order, the solution is that of an unwarped disc.  For the
vertical equilibrium, equation (\ref{f4}) at $O(1)$ yields simply
\begin{equation}
f_{20}=1.
\end{equation}
Also
\begin{equation}
f_{60}=1,
\end{equation}
since $\langle f_6\rangle=1$ by definition.

\subsection{First-order solution}

The horizontal velocities at first order are determined by equation
(\ref{f3}) at $O(|\psi|)$,
\begin{equation}
f_{31}^\prime(\phi)+\alpha
f_{31}(\phi)-2f_{51}(\phi)=\cos\phi-\alpha\sin\phi,
\end{equation}
and equation (\ref{f5}) at $O(|\psi|)$,
\begin{equation}
f_{51}^\prime(\phi)+\alpha
f_{51}(\phi)+{\textstyle{{1}\over{2}}}\tilde\kappa^2f_{31}(\phi)=
{\textstyle{{1}\over{2}}}(4-\tilde\kappa^2)\alpha\cos\phi.
\end{equation}
The solution is of the form
\begin{eqnarray}
f_{31}(\phi)&=&C_{r1}\cos\phi+S_{r1}\sin\phi,\\
f_{51}(\phi)&=&C_{\phi1}\cos\phi+S_{\phi1}\sin\phi,
\end{eqnarray}
with
\begin{equation}
\left[\matrix{\alpha&1&-2&0\cr -1&\alpha&0&-2\cr
{\textstyle{{1}\over{2}}}\tilde\kappa^2&0&\alpha&1\cr
0&{\textstyle{{1}\over{2}}}\tilde\kappa^2&-1&\alpha\cr}\right]\left[
\matrix{C_{r1}\cr
S_{r1}\cr C_{\phi1}\cr S_{\phi1}\cr}\right]=\left[\matrix{1\cr
-\alpha\cr {\textstyle{{1}\over{2}}}(4-\tilde\kappa^2)\alpha\cr
0\cr}\right].
\end{equation}
This may be expressed more compactly using a complex notation, with
(generically)
\begin{equation}
Z=C+{\rm i}S,
\end{equation}
so that
\begin{equation}
\left[\matrix{\alpha-{\rm i}&-2\cr
{\textstyle{{1}\over{2}}}\tilde\kappa^2&\alpha-{\rm
i}\cr}\right]\left[\matrix{Z_{r1}\cr
Z_{\phi1}\cr}\right]=\left[\matrix{1-{\rm i}\alpha\cr
{\textstyle{{1}\over{2}}}(4-\tilde\kappa^2)\alpha\cr}\right].\label{zrzp}
\end{equation}
The determinant of this matrix is
\begin{equation}
-(1-\tilde\kappa^2)-2{\rm i}\alpha+\alpha^2,
\end{equation}
and so a solution exists unless the disc is both Keplerian
($\tilde\kappa^2=1$) and inviscid ($\alpha=0$) to the accuracy of
these scalings.  (As anticipated, the resonant case cannot be
described using the present expansions.)  The solution follows by
inversion of the matrix.  In detail,
\begin{eqnarray}
Z_{r1}&=&{{{\rm i}-(4-\tilde\kappa^2)\alpha+{\rm
i}\alpha^2}\over{(1-\tilde\kappa^2)+2{\rm i}\alpha-\alpha^2}},\\
Z_{\phi1}&=&{{1}\over{2}}\left[{{\tilde\kappa^2+2{\rm
i}(2-\tilde\kappa^2)\alpha-(4-\tilde\kappa^2)\alpha^2}\over
{(1-\tilde\kappa^2)+2{\rm
i}\alpha-\alpha^2}}\right].
\end{eqnarray}

\subsection{Second-order solution}

The enthalpy and vertical velocity at second order are determined by
equation (\ref{f1}) at $O(|\psi|^2)$,
\begin{equation}
f_{12}^\prime(\phi)=(\Gamma-1)f_{42}(\phi)f_{10},
\end{equation}
equation (\ref{f2}) at $O(|\psi|^2)$,
\begin{equation}
f_{22}^\prime(\phi)=(\Gamma+1)f_{42}(\phi),
\end{equation}
and equation (\ref{f4}) at $O(|\psi|^2)$,
\begin{equation}
f_{42}^\prime(\phi)+(\alpha_{\rm
b}+{\textstyle{{4}\over{3}}}\alpha)f_{42}(\phi)=-f_{31}^\prime(\phi)
\cos\phi+(2\sin\phi-\alpha\cos\phi)f_{31}(\phi)-f_{22}(\phi)+
\alpha\cos\phi\sin\phi.
\end{equation}
These may be combined to give
\begin{equation}
f_{42}^{\prime\prime}(\phi)+(\alpha_{\rm
b}+{\textstyle{{4}\over{3}}}\alpha)f_{42}^\prime(\phi)+
(\Gamma+1)f_{42}(\phi)=(3C_{r1}-\alpha
S_{r1}+\alpha)\cos2\phi+(3S_{r1}+\alpha C_{r1})\sin2\phi.
\end{equation}
The solution is of the form
\begin{equation}
f_{42}(\phi)=C_{\theta2}\cos2\phi+S_{\theta2}\sin2\phi,
\end{equation}
with
\begin{equation}
\left[\matrix{-(3-\Gamma)&2(\alpha_{\rm
b}+{\textstyle{{4}\over{3}}}\alpha)\cr -2(\alpha_{\rm
b}+{\textstyle{{4}\over{3}}}\alpha)&-(3-\Gamma)\cr}\right]\left[
\matrix{C_{\theta2}\cr
S_{\theta2}\cr}\right]=\left[\matrix{3C_{r1}-\alpha S_{r1}+\alpha\cr
3S_{r1}+\alpha C_{r1}\cr}\right].
\end{equation}
In a complex notation, the solution is
\begin{eqnarray}
Z_{\theta2}&=&{{-(3+{\rm i}\alpha)Z_{r1}-\alpha}\over{(3-\Gamma)+2{\rm
i}(\alpha_{\rm b}+{\textstyle{{4}\over{3}}}\alpha)}}\nonumber\\
&=&{{-3{\rm i}+2(6-\tilde\kappa^2)\alpha-(1+\tilde\kappa^2){\rm
i}\alpha^2+2\alpha^3}\over{\left[(3-\Gamma)+2{\rm i}(\alpha_{\rm
b}+{\textstyle{{4}\over{3}}}\alpha)\right]\left[(1-\tilde\kappa^2)+2{\rm
i}\alpha-\alpha^2\right]}}.
\end{eqnarray}
It then follows that
\begin{eqnarray}
f_{12}(\phi)&=&-{\textstyle{{1}\over{2}}}(\Gamma-1)f_{10}(S_{\theta2}
\cos2\phi-C_{\theta2}\sin2\phi),\\
f_{22}(\phi)&=&-{\textstyle{{1}\over{2}}}(\Gamma+1)(S_{\theta2}\cos2\phi-
C_{\theta2}\sin2\phi)+{\textstyle{{1}\over{2}}}S_{r1}-
{\textstyle{{1}\over{2}}}\alpha
C_{r1},\\ f_{62}(\phi)&=&S_{\theta2}\cos2\phi-C_{\theta2}\sin2\phi.
\end{eqnarray}

\subsection{Third-order solution}

The horizontal velocities at third order are determined by equation
(\ref{f3}) at $O(|\psi|^3)$,
\begin{eqnarray}
\lefteqn{f_{33}^\prime(\phi)+\alpha
f_{33}(\phi)-2f_{53}(\phi)=f_{42}(\phi)f_{31}(\phi)+\left[f_{22}(\phi)+
(\alpha_{\rm
b}+{\textstyle{{1}\over{3}}}\alpha)f_{42}(\phi)\right]\cos\phi-
\alpha\left[f_{22}(\phi)+\cos^2\phi\right]f_{31}(\phi)}&\nonumber\\
&&\qquad-\alpha\sin\phi\,f_{22}(\phi),
\end{eqnarray}
and equation (\ref{f5}) at $O(|\psi|^3)$,
\begin{equation}
f_{53}^\prime(\phi)+\alpha
f_{53}(\phi)+{\textstyle{{1}\over{2}}}\tilde\kappa^2f_{33}(\phi)=
f_{42}(\phi)f_{51}(\phi)-\alpha\left[f_{22}(\phi)+\cos^2\phi\right]
f_{51}(\phi)+{\textstyle{{1}\over{2}}}(4-\tilde\kappa^2)\alpha\cos\phi\,
f_{22}(\phi).
\end{equation}
The solution is of the form
\begin{eqnarray}
f_{33}(\phi)&=&C_{r3}\cos\phi+S_{r3}\sin\phi+\{\hbox{$m=3$ terms}\},\\
f_{53}(\phi)&=&C_{\phi3}\cos\phi+S_{\phi3}\sin\phi+\{\hbox{$m=3$ terms}\},
\end{eqnarray}
with
\begin{equation}
\left[\matrix{\alpha-{\rm i}&-2\cr
{\textstyle{{1}\over{2}}}\tilde\kappa^2&\alpha-{\rm
i}\cr}\right]\left[\matrix{Z_{r3}\cr
Z_{\phi3}\cr}\right]=\left[\matrix{R_1\cr R_2\cr}\right],
\end{equation}
where
\begin{eqnarray}
R_1&=&\left[{\textstyle{{1}\over{2}}}-{\textstyle{{1}\over{4}}}(\Gamma+1)
{\rm
i}\alpha\right]Z_{\theta2}Z_{r1}^*+\left[{\textstyle{{1}\over{4}}}
(\Gamma+1)({\rm
i}-\alpha)+{\textstyle{{1}\over{2}}}(\alpha_{\rm
b}+{\textstyle{{1}\over{3}}}\alpha)\right]Z_{\theta2}-
{\textstyle{{3}\over{4}}}\alpha
Z_{r1}-{\textstyle{{1}\over{2}}}(S_{r1}-\alpha C_{r1})\left[\alpha
Z_{r1}-(1-{\rm i}\alpha)\right]\nonumber\\
&&\qquad+{\textstyle{{1}\over{2}}}{\rm i}\alpha S_{r1},\\
R_2&=&\left[{\textstyle{{1}\over{2}}}-{\textstyle{{1}\over{4}}}
(\Gamma+1){\rm
i}\alpha\right]Z_{\theta2}Z_{\phi1}^*+{\textstyle{{1}\over{8}}}
(4-\tilde\kappa^2)(\Gamma+1){\rm
i}\alpha Z_{\theta2}-{\textstyle{{3}\over{4}}}\alpha
Z_{\phi1}-{\textstyle{{1}\over{2}}}\alpha(S_{r1}-\alpha
C_{r1})\left[Z_{\phi1}-{\textstyle{{1}\over{2}}}(4-\tilde\kappa^2)
\right]+{\textstyle{{1}\over{2}}}{\rm
i}\alpha S_{\phi1}.
\end{eqnarray}
The $m=3$ terms will not be required.

\subsection{Evaluation of the coefficients}

The coefficients $Q_1$ and $Q_4$ are even in $|\psi|$ and have
expansions
\begin{eqnarray}
Q_1&=&Q_{10}+|\psi|^2Q_{12}+O(|\psi|^4),\\
Q_4&=&Q_{40}+|\psi|^2Q_{42}+O(|\psi|^4).
\end{eqnarray}
At leading order,
\begin{equation}
Q_{10}=-{\textstyle{{1}\over{2}}}(4-\tilde\kappa^2)\alpha
\end{equation}
and
\begin{eqnarray}
Q_{40}&=&\big\langle{\rm e}^{{\rm i}\phi}\left[(1-{\rm
i}\alpha)f_{31}(\phi)-{\rm
i}\alpha\sin\phi\right]\big\rangle\nonumber\\
&=&{\textstyle{{1}\over{2}}}\alpha+{\textstyle{{1}\over{2}}}(1-{\rm
i}\alpha)Z_{r1}\nonumber\\ &=&{{1}\over{2}}\left[{{{\rm
i}-2\alpha+(7-\tilde\kappa^2){\rm
i}\alpha^2}\over{(1-\tilde\kappa^2)+2{\rm i}\alpha-\alpha^2}}\right].
\end{eqnarray}
At second order,
\begin{eqnarray}
Q_{12}&=&-{\textstyle{{1}\over{4}}}(4-\tilde\kappa^2)\alpha(S_{r1}-\alpha
C_{r1})-{\textstyle{{1}\over{2}}}(C_{r1}C_{\phi1}+S_{r1}S_{\phi1})+
{\textstyle{{1}\over{2}}}\alpha C_{\phi1}\nonumber\\
&=&{{1}\over{4}}\left[{{-(8-12\tilde\kappa^2+3\tilde\kappa^4)\alpha-
(2-\tilde\kappa^2)(28-12\tilde\kappa^2+\tilde\kappa^4)\alpha^3+
(8-\tilde\kappa^2)(4-\tilde\kappa^2)\alpha^5}\over{(1-\tilde\kappa^2)^2+
2\alpha^2(1+\tilde\kappa^2)+\alpha^4}}\right]
\end{eqnarray}
and
\begin{eqnarray}
Q_{42}&=&\big\langle{\rm e}^{{\rm i}\phi}\left\{(1-{\rm
i}\alpha)f_{33}(\phi)-{\rm
i}\left[f_{31}(\phi)-\alpha\cos\phi\right]\left[f_{42}(\phi)+
f_{31}(\phi)\cos\phi\right]-{\rm
i}\alpha f_{22}(\phi)f_{31}(\phi)-{\rm i}\alpha
f_{22}(\phi)\sin\phi\nonumber\right.\\
&&\left.\qquad+f_{62}(\phi)\left[(1-{\rm i}\alpha)f_{31}(\phi)-{\rm
i}\alpha\sin\phi\right]\right\}\big\rangle\nonumber\\
&=&{\textstyle{{1}\over{2}}}(1-{\rm
i}\alpha)Z_{r3}+{\textstyle{{1}\over{2}}}\left[-{\rm
i}+{\textstyle{{1}\over{4}}}(\Gamma-1)\alpha\right]Z_{\theta2}Z_{r1}^*+
{\textstyle{{1}\over{8}}}(3-\Gamma){\rm
i}\alpha Z_{\theta2}-{\textstyle{{1}\over{4}}}{\rm
i}Z_{r1}Z_{r1}^*-{\textstyle{{1}\over{8}}}{\rm
i}Z_{r1}^2+{\textstyle{{3}\over{8}}}{\rm i}\alpha Z_{r1}\nonumber\\
&&\qquad-{\textstyle{{1}\over{4}}}{\rm i}\alpha(S_{r1}-\alpha
C_{r1})(Z_{r1}+{\rm i})+{\textstyle{{1}\over{4}}}\alpha
S_{r1}\nonumber\\ &=&{{a+b\Gamma+c\left[\Gamma-2{\rm i}(\alpha_{\rm
b}+{\textstyle{{4}\over{3}}}\alpha)\right]}\over{4\left[(3-\Gamma)+2{\rm
i}(\alpha_{\rm
b}+{\textstyle{{4}\over{3}}}\alpha)\right]\left[(1-\tilde\kappa^2)+2{\rm
i}\alpha-\alpha^2\right]^3\left[(1-\tilde\kappa^2)-2{\rm
i}\alpha-\alpha^2\right]}},
\end{eqnarray}
where
\begin{eqnarray}
a&=&6{\rm
i}(1-\tilde\kappa^2)^2-3(1-\tilde\kappa^2)(8-7\tilde\kappa^2+\tilde\kappa^4)
\alpha+3(1-\tilde\kappa^2)(28-32\tilde\kappa^2+2\tilde\kappa^4+
\tilde\kappa^6){\rm i}\alpha^2\nonumber\\
&&\qquad-(258-337\tilde\kappa^2+103\tilde\kappa^4-15\tilde\kappa^6+
3\tilde\kappa^8)\alpha^3+(224-750\tilde\kappa^2+511\tilde\kappa^4-
106\tilde\kappa^6+6\tilde\kappa^8){\rm i}\alpha^4\nonumber\\
&&\qquad-(442-511\tilde\kappa^2+116\tilde\kappa^4+2\tilde\kappa^6)
\alpha^5-(552-560\tilde\kappa^2+124\tilde\kappa^4-7\tilde\kappa^6){\rm
i}\alpha^6+(562-153\tilde\kappa^2-\tilde\kappa^4)\alpha^7\nonumber\\
&&\qquad+(66-26\tilde\kappa^2+\tilde\kappa^4){\rm
i}\alpha^8+2(1-2\tilde\kappa^2)\alpha^9-4{\rm i}\alpha^{10},\\
b&=&-(4-\tilde\kappa^2)\alpha\left[(1-\tilde\kappa^2)+2{\rm
i}\alpha-\alpha^2\right]\left[(1-\tilde\kappa^2)-(2-\tilde\kappa^2){\rm
i}\alpha+\alpha^2\right]\left[3+2(6-\tilde\kappa^2){\rm
i}\alpha+(1+\tilde\kappa^2)\alpha^2+2{\rm i}\alpha^3\right],\\
c&=&{\rm
i}(1-\tilde\kappa^2)^2-2(2-\tilde\kappa^2)(1-\tilde\kappa^2)
(1+\tilde\kappa^2)\alpha-(2-\tilde\kappa^2)(1-\tilde\kappa^2)
(2+9\tilde\kappa^2-3\tilde\kappa^4){\rm
i}\alpha^2-2(25-40\tilde\kappa^2+24\tilde\kappa^4-3\tilde\kappa^6)
\alpha^3\nonumber\\
&&\qquad-(262-428\tilde\kappa^2+233\tilde\kappa^4-44\tilde\kappa^6+
2\tilde\kappa^8){\rm
i}\alpha^4+2(115-119\tilde\kappa^2+23\tilde\kappa^4-\tilde\kappa^6)
\alpha^5\nonumber\\
&&\qquad+(164-196\tilde\kappa^2+45\tilde\kappa^4-2\tilde\kappa^6){\rm
i}\alpha^6-2(87-18\tilde\kappa^2+\tilde\kappa^4)\alpha^7-3
(9-2\tilde\kappa^2){\rm
i}\alpha^8-2\alpha^9.
\end{eqnarray}

\label{lastpage}

\end{document}